\documentclass[10pt,journal,letterpaper,compsoc]{IEEEtran}

\usepackage{cite}
\usepackage{url}
\usepackage[pdftex]{graphicx}
\usepackage[cmex10]{amsmath}
\usepackage{array}
\usepackage[caption=false,font=footnotesize,labelfont=sf,textfont=sf]{subfig}
\usepackage{fixltx2e}
\usepackage{stfloats}
\usepackage{multirow}
\usepackage{color}
\usepackage{amssymb}
\usepackage[english]{babel}
\usepackage{amsthm}

\usepackage[lined,vlined,ruled,commentsnumbered]{algorithm2e}

\newtheorem{theorem}{Theorem}

\newtheorem{proposition}{Proposition}

\ifodd 1211
\newcommand{\revaa}[1]{{#1}}
\newcommand{\revxx}[1]{{#1}}

\newcommand{\revv}[1]{{#1}}
\newcommand{\comm}[1]{\textbf{\color{red} (Comment: #1) }}

\newcommand{\revh}[1]{#1}
\newcommand{\com}[1]{\textbf{\color{red} (Comment: #1) }}
\newcommand{\comg}[1]{\textbf{\color{green} (COMMENT: #1)}}
\newcommand{\response}[1]{\textbf{\color{green} (RESPONSE: #1)}}
\else
\newcommand{\revaa}[1]{#1}
\newcommand{\revxx}[1]{#1}

\newcommand{\revv}[1]{#1}
\newcommand{\comm}[1]{}

\newcommand{\revh}[1]{#1}
\newcommand{\com}[1]{}
\newcommand{\comg}[1]{}
\newcommand{\response}[1]{}
\fi

\newcommand{\revx}[1]{{#1}}

\newcommand{\scfootnote}[1]{\footnote{#1}}

\def\dd{\mathrm{d}}
\def\eq{\triangleq}

\def\sumN{\sum\limits_{n=1}^N}
\def\sumM{\sum\limits_{m=1}^N}

\def\sumTa{\sum\limits_{\t=1}^T}
\def\sumZ{\sum\limits_{z=1}^Z}

\def\T{\mathcal{T}}
\def\N{\mathcal{N}}
\def\L{\mathcal{A}}

\def\tseg{\beta}
\def\btseg{\boldsymbol{\tseg}}
\def\R{\mathcal{R}}
\def\Q{Q}

\def\dset{\boldsymbol{S}}
\def\ds{\boldsymbol{s}}

\def\du{u}
\def\dr{r}
\def\dz{z}
\def\dts{t^{\mathrm{s}}}
\def\dte{t^{\mathrm{e}}}
\def\dunk{\du_{n[k]}}
\def\drnk{\dr_{n[k]}}
\def\dznk{\dz_{n[k]}}
\def\dtsnk{\dts_{n[k]}}
\def\dtenk{\dte_{n[k]}}

\def\rset{\widehat{\boldsymbol{S}}}
\def\rs{\hat{\boldsymbol{s}}}

\def\ru{\hat{u}}
\def\rr{\hat{r}}
\def\rz{\hat{z}}
\def\rts{\hat{t}^{\mathrm{s}}}
\def\rte{\hat{t}^{\mathrm{e}}}
\def\runk{\ru_{n[k]}}
\def\rrnk{\rr_{n[k]}}

\def\rtenk{\rte_{n[k]}}
\def\rumk{\ru_{m[k]}}
\def\rrmk{\rr_{m[k]}}
\def\rzmk{\rz_{m[k]}}
\def\rtsmk{\rts_{m[k]}}
\def\rtemk{\rte_{m[k]}}

\def\buf{q}
\def\h{h}

\def\SW{W}
\def\Pay{P}
\def\Ut{U}
\def\Va{V} 
\def\va{\theta}
\def\Lossdeg{L^{\textsc{qdeg}}}	
\def\ld{\phi^{\textsc{qdeg}}}
\def\Lossrebuf{L^{\textsc{rebuf}}}	
\def\lr{\phi^{\textsc{rebuf}}}

\def\Cost{C}
\def\Costc{E^{\textsc{cell}}}
\def\Costw{E^{\textsc{wifi}}}

\def\cct{c^{\textsc{time}}}
\def\ccv{c^{\textsc{data}}}
\def\cwt{w^{\textsc{time}}}
\def\cwv{w^{\textsc{data}}}

\def\xSW{\widetilde{\SW}}
\def\xPay{\widetilde{\Pay}}

\def\xVa{\widetilde{\Va}} 
\def\xLossdeg{\widetilde{L}^{\textsc{qdeg}}}	
\def\xLossrebuf{\widetilde{L}^{\textsc{rebuf}}}	

\def\xCostc{\widetilde{E}^{\textsc{cell}}}
\def\xCostw{\widetilde{E}^{\textsc{wifi}}}

\def\SWo{\SW^{*}}
\def\SWx{\xSW^{*}}
\def\SWi{\SW^{\prime}}

\def\t{\tau}

\def\kvec{\boldsymbol{K}}		
\def\k{\kappa}
\def\x{x}		
\def\xd{\x^{\textsc{dl}}}
\def\y{y}		
\def\yr{\y^{\textsc{re}}}

\begin{document}

\title{\vspace{-5mm}
Multi-User Cooperative Mobile Video Streaming: Performance Analysis and Online Mechanism Design}

\author{	\vspace{-2mm}
Lin~Gao, \emph{Senior Member, IEEE},
Ming~Tang,
Haitian~Pang,  \emph{Student Member, IEEE},
Jianwei~Huang,  \emph{Fellow, IEEE},
and Lifeng~Sun, \emph{Member, IEEE}
		\vspace{-2mm}
\IEEEcompsocitemizethanks{
\IEEEcompsocthanksitem
L.~Gao is with the School of Electronic and Information Engineering, Harbin Institute of Technology, Shenzhen, China, E-mail: gaol@hit.edu.cn;
M.~Tang and J.~Huang are with the Network Communications and Economics Lab (NCEL),
Department of Information Engineering, The Chinese University of Hong Kong, E-mail: \{mtang, jwhuang\}@ie.cuhk.edu.hk;
H.~Pang and L.~Sun are with the Department of Computer Science and Technology, Tsinghua University, China, E-mail:
pht14@mails.tsinghua.edu.cn, sunlf@tsinghua.edu.cn.
}}

\IEEEcompsoctitleabstractindextext{%
\begin{abstract}
\revxx{Adaptive bitrate streaming enables video users to \emph{adapt} their playing bitrates to the real-time network conditions, hence achieving the desirable quality-of-experience (QoE).
In a multi-user wireless scenario, however, existing single-user based bitrate adaptation methods may fail to provide the desirable QoE, due to lack of consideration of multi-user interactions (such as the multi-user interferences and network congestion).
In this work, we propose a novel user cooperation framework based on \emph{user-provided networking} for multi-user mobile video streaming over wireless cellular networks.
The framework enables nearby mobile video users to crowdsource their cellular links and resources for cooperative video streaming.
We first analyze the social welfare performance bound of the proposed cooperative streaming system by introducing a virtual time-slotted system.
Then, we design a low complexity Lyapunov-based online algorithm,  which can be implemented in an online and distributed manner without the complete future and global network information.
Numerical results show that the proposed online algorithm achieves an average $97\%$ of the theoretical maximum social welfare.
We further conduct experiments with real data traces, to compare our proposed online algorithm with the existing online algorithms in the literature.
Experiment results show that our algorithm outperforms the existing algorithms in terms of both the achievable bitrate (with an average gain of $20\%\sim30\%$)
and social welfare (with an average gain of $10\%\sim50\%$).}
\vspace{-2mm}
\end{abstract}

\begin{IEEEkeywords}
Mobile Video Streaming,
Adaptive Bitrate,
Mobile Crowdsourcing,
Online Algorithm
\end{IEEEkeywords}
}

\maketitle

\IEEEdisplaynotcompsoctitleabstractindextext

\IEEEpeerreviewmaketitle

\addtolength{\abovedisplayskip}{-1mm}
\addtolength{\belowdisplayskip}{-1mm}


\section{Introduction}

\subsection{Background and Motivations}


%
%
%

Global mobile data traffic is growing at an unprecedented rate, where mobile video streaming contributes most of the data growth.
According to  Cisco \cite{CiscoReport}, mobile video streaming traffic has accounted for 60\% of the global mobile data traffic in 2016, and the percentage is expected to   increase to 78\% by 2021.
\emph{Adaptive BitRate (ABR)} streaming \cite{abr} is a promising technology for video streaming over large distributed HTTP networks (e.g., Internet) and has been adopted by many popular online video streaming systems (e.g.,
HTTP dynamic streaming of Adobe \cite{adobe}, HTTP live streaming of Apple  \cite{apple},
and smooth streaming of Microsoft~\cite{microsoft}).
The key idea of ABR is to enable video players to \emph{adapt} the playing bitrate (corresponding to the quality of video, e.g., resolution) to the real-time network conditions to ensure the desirable quality-of-experience (QoE).
While most of the existing works focused on the bitrate adaptation methods of a \emph{single user} (e.g., \cite{b1,b6}), in this work we consider a more general scenario of \emph{multi-user} video streaming over wireless cellular networks.
In the multi-user wireless scenario, the QoE of each mobile   user is affected not only by the stochastic changing
of his own network condition (e.g., channel fading),
but also by the potential resource competition and interference of
other users \cite{ming-maga, c1x, c2, c3,   c6, multiuser-dash-1, ming-2017, ming-2016a, ming-2016b, add-1, add-3,add-3a}.
Without proper coordination or cooperation among users, such competition and interference may degrade the network performance greatly (e.g., leading to congestion),
hence increase the video streaming cost
and harm the user QoE.
Thus, the existing single-user based bitrate adaptation methods often fail to provide desirable QoE for video users in the multi-user scenario,
due to lack of consideration of multi-user competition and interference.

\begin{figure}[t]
\vspace{-3mm}
  \centering
~~~~\includegraphics[height=1.3in]{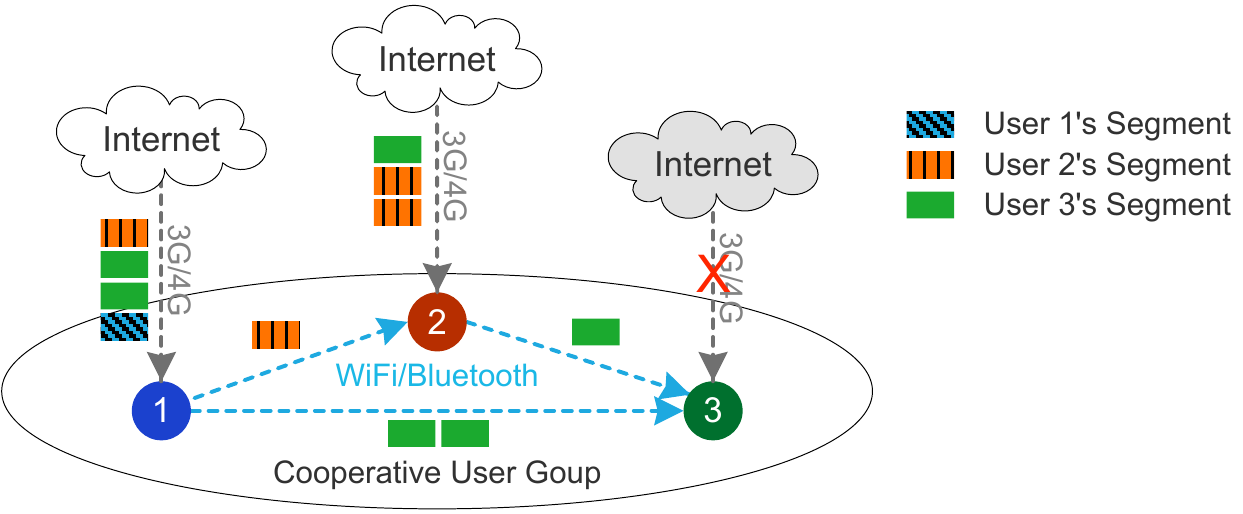}
  \caption{Cooperative  Video Streaming Model.}
  \label{fig:model}
  \vspace{-4mm}
\end{figure}

To this end, in this work we will study the multi-user \emph{cooperative} video streaming, where (nearby) mobile video users cooperate with each other in both bitrate adapting and video downloading.
Namely, each user can download video data for other users using his own cellular link or download his video data through others' links.
In this sense, users \emph{aggregate} their cellular links and resources for the cooperative video streaming.
Figure \ref{fig:model} illustrates such a cooperative streaming model with three users \{1, 2, 3\}, where user 1 downloads for all three users and user 2 downloads for himself and user 3.
Note that user 3 does not have the available cellular link. ~~~~~~~~~~~~~~~~~~

\revaa{There are several real-world scenarios where the multi-user cooperative streaming is useful and helpful.
First, the most relevant scenario is \emph{User-Provided Networking} (UPN) \cite{upn-lin,opengarden-lin}, where mobile devices (e.g., 4G smartphones) with abundant cellular link capacities operate as mobile hotspots and provide Internet access to other devices.
UPN has been widely studied and implemented today, and some IT and Telecom companies (such as OpenGarden \cite{opengarden}, Karma \cite{karma}, and AT\&T \cite{att}) have provided commercial UPN services.
{The cooperative streaming proposed in this work can enhance the capability of UPN on providing the video streaming service.}
Another more concrete scenario is \emph{Mobile Live Streaming} (MLS) \cite{live-1,live-2}, with which people can watch live activities of their friends or share their own activities to their friends on their smartphones.
MLS becomes popular in recent years with the proliferation of smartphones and 4G cellular networks.
Nowadays, many social network companies have provided MLS services, such as IngKee \cite{liveapp-1}, Youtube Live \cite{liveapp-2}, and Facebook Livestream \cite{liveapp-3}.
{The cooperative streaming proposed in this work can improve the live streaming quality in MLS.}}

\revaa{The key motivation for considering such a cooperative streaming system is the \emph{heterogeneity} of mobile devices.\footnote{\revaa{According to   \cite{CiscoReport}, smartphones only account  for 38\% of the total mobile devices, and a large amount of non-smartphone mobile devices (e.g., tablets and laptops) still lack   stable and always-on Internet connections, especially in the outdoor environment. Our proposed system can   help
these devices   connect to the Internet via the links of nearby smartphones.
Furthermore, even for the smartphone devices with the same or similar Internet capability, they
may have different cost evaluations for energy consumption (depending on, for example, their battery status), resulting in certain ``heterogeneity''
among devices.}}
Note that cooperative streaming can be easily implemented in a practical scenario (such as UPN and MLS) by installing some customized apps (e.g., OpenGarden \cite{opengarden}) on smartphones, and the related optimization and incentive issues have been studied in the recent literature (e.g., \cite{upn-lin, opengarden-lin}).
\revx{However, the existing techniques in \cite{upn-lin, opengarden-lin} cannot be directly applied to the cooperative video streaming model,
 due to the asynchronous operations of video streaming and the unique QoE requirements of video applications.
This motivates us to study the multi-user cooperative streaming in this work.}}

\ifodd 2
Motivations for such a cooperative streaming model are as follows.
\revx{First, the majority of today's mobile users (e.g., $74\%$ in 2016 \cite{CiscoReport}) are still using 2G/3G cellular systems and devices, which can only provide very low link capacities and cannot support high quality video streaming.
The cooperative streaming model enables  users with high cellular link capacities (e.g., those using 4G systems/devices) to download video for users with poor cellular link capacities (e.g., those using 2G/3G systems/devices) or even without available cellular links (e.g., those using tablets or laptops).}
Second, by exploiting the user diversity of resource availability and   service requirement, this cooperative streaming model can potentially  reduce the negative network  externality (e.g., competition and interference among users), while amplifying  the positive network effect (e.g., cooperation among users).\footnote{\revx{Specifically, without cooperation, users download their videos separately, hence
can potentially compete/interfere with each other (for the wireless access resource or the video server resource) when streaming videos through the same wireless access network (such as the same cellular base station).
With cooperation, however, users are scheduled to download video streams jointly and cooperatively, hence can potentially avoid competition and interference.}}
Moreover, such a cooperative streaming framework
can be easily implemented in practice by installing some customized mobile apps on smartphones (e.g., OpenGarden \cite{opengarden}), and the related optimization and incentive issues have been studied in the recent literature (e.g., \cite{upn-lin,opengarden-lin}).
\revx{Nevertheless, the techniques in \cite{upn-lin,opengarden-lin} cannot be directly applied to the video streaming model,
 due to the asynchronous operations in video streaming and the unique QoE requirements of video applications.
This motivates us to study the multi-user cooperative streaming in this work.}
\fi

 \vspace{-3mm}

\subsection{Solution and Contributions}

%
%
%
%

In this work, we propose a general \emph{multi-user cooperative video streaming} framework based on UPN \cite{upn-lin,opengarden-lin}.
The key idea is to enable nearby mobile users to form a cooperative group (via WiFi or Bluetooth) and aggregate their cellular links and resources for the cooperative video downloading and bitrate adapting.
We focus on studying the users' streaming behaviours (i.e., \emph{download scheduling} and \emph{bitrate adaptation}) in the proposed cooperative  framework.
Namely, for each video user, when and from whom he is going to download each video segment, at which bitrate?
Our goal is to understand the performance bound of the system and design an online scheduling method to approach such a bound.



\vspace{1mm}

First, \emph{we formally define the users' operations in the cooperative  streaming system, and formulate the corresponding  social welfare optimization problem  (Section \ref{sec:formulation}).}
The optimal solution of this problem provides the theoretical performance bound of the proposed cooperative streaming system. \revx{A comprehensive analysis for such a performance bound is the foundation of the future study on privacy, security, and incentive mechanism design.}\footnote{
{In practice, there have been various existing approaches to address the privacy/security issue in the similar systems.
For example, Opengarden \cite{opengarden} solves the privacy/security issue by using
advanced data encryption technique through built-in softwares, while
Karma \cite{karma} solves it by using hardware-enabled
data encryption technique through dedicated devices.
These existing technical ways can
help us quickly build the privacy and security system.}
}

\vspace{1mm}

\revxx{Second, \emph{we analyze the social welfare performance bound of the proposed cooperative  streaming system   (Section \ref{sec:main1}).}}
Directly solving such a performance
is challenging, due to the asynchronous operations of users as well as the mixed-integer nature of the problem.
To this end, we introduce a \emph{virtual} time-slotted system with the synchronized operations, and formulate the  new social welfare optimization problem as a linear programming (which can be solved efficiently with many standard methods).
We show that with proper choices of time parameters, the optimal solution of the virtual time-slotted system can provide an effective upper-bound and lower-bound for the optimal solution (performance bound) of the original   system, which forms the feasible performance region of the proposed cooperative streaming system.

\vspace{1mm}

\revxx{Finally, \emph{we design a Lyapunov-based online streaming algorithm  for the practical implementation of the proposed cooperative streaming system (Section \ref{sec:main2}).}
The proposed algorithm converges to the theoretical performance bound asymptotically, with a controllable approximation error bound.}
Moreover, it relies only on the current state and historical streaming information (while not on any future network information), hence can be implemented in the \emph{online} manner; and it requires only the local information exchange within each cooperative group (while not the global network information exchange), hence can be implemented in the \emph{distributed} manner.
We perform extensive experimental simulations with real data traces to evaluate its performance gap with the theoretical bound and to compare its performance with state-of-art online algorithms in the existing literature.

\vspace{1mm}

For more clarity, we summarize the logical relationship among the above three parts as follows:
(i) \emph{the social welfare optimization problem in Section \ref{sec:formulation} defines the theoretical performance bound of the proposed cooperative streaming system (but it is challenging to solve);}
(ii) \emph{the virtual time-slotted system in Section \ref{sec:main1} helps    characterize the region (i.e., upper-bound and lower-bound) of the above theoretical performance bound;}
(iii) \emph{the online algorithm in Section \ref{sec:main2} converges to the above theoretical performance bound asymptotically in the realistic scenario without complete future network information.}
More specifically, the key contributions of this work are summarized as follows.~~~~~~~~~~~~~~~~~~~~~~~~~~~~~

\begin{itemize}

\item \emph{Novel Model:}
To our best knowledge, this is the first work that proposes a general multi-user cooperative streaming framework for mobile video streaming.
\revv{The framework enables mobile video users to crowdsource their radio connections and resources for cooperative video streaming,
and can effectively improve the QoE of video users.}
Moreover, we provide  both theoretical performance analysis and practical algorithm design for such a cooperative streaming system.


\vspace{1mm}

\item \emph{Performance Bound Analysis:}
We analyze the theoretical performance bound of the proposed cooperative streaming system,
overcoming the challenging issue of asynchronous operations by using a virtual time-slotted system.
Such a performance bound analysis is
fundamental for the design, evaluation, and implementation of practical algorithms in such a cooperative streaming system.


\vspace{1mm}

\item \emph{Online Algorithm Design:}
We implement the cooperative streaming system in the practical scenario without future and global network information, and design a Lyapunov-based online streaming algorithm.
The proposed algorithm converges to the theoretical performance bound asymptotically. ~~~~~~

\vspace{1mm}

\item \emph{Experiment and Demo:}
We conduct extensive experiments with real data traces, which show that our proposed cooperative streaming system, together with the online streaming  algorithm, outperforms the existing systems and algorithms in terms of both achieved bitrate (with an average gain of $20\%\sim30\%$) and social welfare (with an average gain of $10\%\sim50\%$).
\revx{We also construct a real demo system to implement and evaluate the proposed system and algorithm.}

\end{itemize}

\vspace{1mm}


The rest of the paper is organized as follows.
In Section \ref{sec:literature}, we review the related work.
In Section \ref{sec:model}, we present the system model.
In Section \ref{sec:formulation}, we provide the problem formulation.
In Section \ref{sec:main1}, we propose the virtual time-slotted system and the performance bound analysis.
In Section \ref{sec:main2}, we propose the Lyapunov-based online streaming algorithm.
We provide simulation results in Section \ref{sec:simulation} and conclude in Section \ref{sec:con}.


\section{Literature Review}
\label{sec:literature}

Prior works on ABR video streaming mainly focused on the bitrate adaptation of a \emph{single user} using either   {buffer-based} method \cite{b1} or   {channel prediction-based}~method \cite{b6}.
Recently, there is a growing interest in exploiting the \emph{multi-user} cooperative video streaming.
From the modeling perspective, the existing cooperative streaming models can be classified into four categories (see \cite{ming-maga} for more details):
  \emph{Bandwidth Aggregation (BA)} model \cite{c6},
 \emph{Device-to-Device (D2D)} model \cite{add-1, add-3,add-3a},
  \emph{Crowdsourced Mobile Streaming (CMS)} model \cite{multiuser-dash-1, ming-2017},
and
  \emph{Mobile Peer-to-Peer (MP2P)} model \cite{c1x,c2,c3}.

\emph{1) BA Model}  \cite{c6}:
The key idea is to aggregate the bandwidth of nearby users to help a particular mobile video user's streaming.
The BA model mainly focused on the simple \emph{one-to-many} cooperation between a single video user and multiple helpers \cite{c6}. 
We consider a more general \emph{many-to-many} cooperation framework with multiple video users and multiple helpers, where each user acts as both the video user and the helper.

\emph{2) D2D Model}  \cite{add-1,  add-3,add-3a}:
The key idea is to enable nearby video users to share their downloaded video segments with each other through D2D links.
In \cite{add-1}, Golrezaei \emph{et al.} studied the cache-based D2D cooperation, where mobile video users cache popular video contents and deliver to other users via D2D links in the future.
Our model differs from that of \cite{add-1} in the following aspects.
First, we consider the real-time cooperation of nearby users, while they considered the future opportunistic cooperation.
Second, we study the jointly video streaming of multiple users, while they studied the video streaming of different users separately. 
In \cite{add-3,add-3a}, researchers studied the real-time D2D based cooperation, where multiple nearby users watch the \emph{same} video and share video contents cooperatively via D2D links.
Our model is similar but more general than those in \cite{add-3,add-3a}, as we allow different users to watch different videos.
This introduces an additional dimension (i.e., video index) when making the scheduling decision, hence involves additional challenges.

\emph{3) CMS Model}  \cite{multiuser-dash-1, ming-2017, ming-2016a, ming-2016b}:
The key idea is to enable nearby mobile video users pool their network resources together to satisfy all users' video streaming requirements jointly. 
Note that our proposed cooperative streaming model falls into this category. 
In \cite{multiuser-dash-1}, Pu \emph{et al.} proposed a rate adaptation algorithm for optimizing the adaptive streaming across multiple mobile users (possibly watching different videos), but they didn't consider the individual characteristics of different users.
In \cite{ming-2017, ming-2016a}, Tang \emph{et al.} focused on the incentive design in the multi-user CMS model and proposed a multi-dimensional auction-based mechanism to incentivize video users to collaborate with each other under information asymmetry. 
However, they neither performed the performance bound analysis, nor designed the online algorithm. 
In \cite{ming-2016b}, Gao \emph{et al.} analyzed the performance bound for multi-user CMS models, but didn't consider the online algorithm design.
In this work, we will study both the theoretical performance bound and the practical online algorithm systematically.




\emph{4) MP2P Model} \cite{c1x, c2, c3}:
The key idea is to enable video users act as virtual video servers and send the downloaded segments to other users via \emph{Internet}.
Thus, in the MP2P model, a video user can potentially help other users that are not physically close-by.
The key difference between our model and the MP2P model is as follows.
In the MP2P model, each video segment has multiple copies residing on
both the video server and the user devices (peers), and video users can download a video segment from either the server or a user peer, via his own wireless cellular link.
Hence, the key design purpose of MP2P model is to \emph{reduce the load of the video server}.
In our cooperative streaming model, however, each video segment has a unique copy residing on the video server, and users can download a video segment (from the video server) either via his own wireless cellular link or a neighbor's cellular link.
Hence, the key design purpose of our model is to \emph{reduce the uncertainty} or \emph{improve the efficiency of user's wireless cellular link}.

%


 \section{System Model}
\label{sec:model}

\subsection{Network Model}


We consider a set $\N \eq \{1,\ldots,N\}$ of mobile video users in wireless cellular networks, who want to watch videos (on their smartphones) via 3G/4G cellular links.
Mobile users are heterogeneous in terms of their cellular link capacities and video quality requirements.
For example, a user requesting a high quality video may suffer from a low cellular link capacity,
due to factors such as a severe channel fading
and a high cellular network congestion.
This may reduce the quality of the video and increase the video quality variation, both harming the user's quality of experience (QoE).
On the other hand, a user requesting a low quality video  {(or not playing a video at all)} may experience a high cellular link capacity, and have extra capacity to help other users.
Thus, it is desirable to enable users to connect with each other to download the streaming video contents cooperatively.

\revx{There are many real-world application scenarios for such a cooperative video streaming. Consider, for example, that a group of friends who want to watch a live soccer match together on their phones   at a remote location (e.g., a camping or skiing site), or a family who wants to watch one or multiple movies on their phones in the train or in the car, or a group of students who want to watch different online lectures using WiFi at a busy hotspot (e.g., a classroom).
In all these cases, some or all of the users may have poor or intermittent cellular connectivity, depending on the coverage of their service providers. Thus, aggregating  the resources of nearby users for the cooperative video streaming may significantly improve the overall user satisfactions.} ~~~~~~~~~~~~~~~~~~~~

\emph{1) \textbf{User-Provided Network (UPN):}}
UPN enables nearby mobile users to form a cooperative group (via WiFi) and aggregate their radio connections and resources for cooperative data downloading.
We consider a general \emph{multi-user cooperative streaming} scheme based on UPN. Namely, in a cooperative group, each user can download video data for other users using his own cellular link (and resources) and download his video data through other users' links (and resources).
As mentioned previously, we assume that some well-designed incentive mechanisms (e.g., auction \cite{inc-1, inc-2, inc-3, inc-4}, contract \cite{cccc-1,cccc-2,cccc-3,cccc-4}, or others trust mechanisms \cite{inc-5, inc-8}) have been adopted, such that users are willing to participate in the cooperative streaming system to help others.

Figure \ref{fig:model} illustrates such a cooperative streaming model with three users \{1, 2, 3\},
where user 1 downloads one segment for himself, one segment for user 2, and two segments for user 3, while user 2 downloads two segments for himself and one segment for user 3.
Note that user 3 does not download any video content due to the temporary interruption of his  cellular link.



\begin{figure}[t]
\vspace{-5mm}
  \centering
  \includegraphics[height=0.9in]{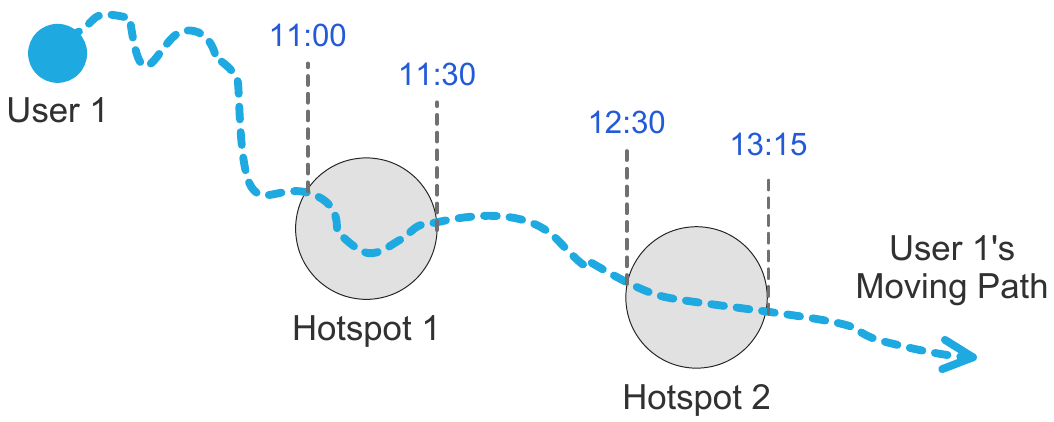}
  \vspace{-2mm}
  \caption{Hotspot-Based Mobility Model.}
  \label{fig:mobility}
  \vspace{-4mm}
\end{figure}

\emph{2) \textbf{Mobility Model:}}
The cooperation gain of such a cooperative streaming highly depends on the number of cooperative users and the duration of cooperation, both closely related to the users' mobility patterns.
We adopt a \emph{hotspot-based} mobility model \cite{mobility},
where the whole area is divided into a set of small hotspots and the non-hotspot area,\footnote{A hotspot is a small area where users are likely to  stay for a substantial amount of  time (e.g., a bus stop or a coffee shop), hence can maintain their WiFi connections  for a reasonable amount  of time.}
and each user moves across a sequence of hotspots during his travel in the following pattern: \emph{staying for a certain period of time in each hotspot that he passes, and taking some time for each transition (from one hotspot to another).}
Figure \ref{fig:mobility} illustrates such a mobility model, where user 1 stays at hotspot 1 for 30 minutes (11:00$\sim$11:30), and then takes 1 hour to move to hotspot 2 and stays at hotspot 2 for 45 minutes  (12:30$\sim$13:15).

In such a hotspot-based mobility model, users in the same hotspot at the same time can connect with each other (hence form a cooperative group), while users in different hotspots or in the non-hotspot area cannot.
Such a mobility model has been widely-used in the scenarios where users need to take certain time to interact with each other (e.g., mobile data forwarding in \cite{mobi-hotspot}).

\emph{\textbf{Notations:}}
We consider the operation in a period of continuous time $\T \eq  [0,\ T]$, where $t = 0$ is the initial time and $T $ is the ending time.
Let $\L \eq \{1, \ldots, A\}$ denote the set of all hotspots, and $\{0\}$ denote the non-hotspot area.
The key notations in this part are listed below.

 \vspace{1mm}
$\bullet$ $a_n(t) \in \L \bigcup \{0\}$: the  location of user $n$ at time $t$;

\vspace{1mm}
$\bullet$ $h_n(t) > 0$: the cellular link capacity of user $n$ at time~$t$;

\vspace{1mm}
$\bullet$ $e_{n,m}(t) \in \{0,1\}$: the indicator denoting whether users $n$ and $m$ are encountered (i.e., in the same hotspot) at time $t$, i.e., $e_{n,m}(t) =1 $ if $a_n(t) = a_m(t) \in \L$.

\vspace{1mm}

For convenience, we refer to the user location and cellular link capacity $\{( a_n(t), h_n(t)), \forall n\in\N,t\in\T \}$ as the \emph{network information}, which   varies randomly over time.
Note that the encounter indicator $e_{n,m}(t)$ can be derived from  the location information  of users $n$ and $m$.


\subsection{Video Streaming Model}




We  consider a typical ABR streaming
model \cite{abr},
 where a single source video file is partitioned into multiple segments and delivered to a video user using HTTP.
The key features of ABR model are summarized below.

(i) \emph{Video Segmenting:}
To facilitate the video delivery over the Internet,
a source video file is divided into a sequence of small HTTP-based file segments, each containing a short interval of playback time (e.g., 2--10 seconds) of the source video, which is possibly several hours in term of the total duration  (e.g., a movie).
A user downloads the video segment by segment.~~~~~~~~~~~~~~~~~~~~~~~~~~~~~~~~~~~~~~

(ii) \emph{Multi-Bitrate Encoding:}
Each segment is encoded at multiple bitrates, each  corresponding to a specific video quality (such as resolution).
A user can select different bitrates for different segments according to real-time network conditions.

(iii)  \emph{Data Buffering:}
For smoothly playing, each downloaded segment is first stored in a buffer at the user's device, and then fetched to the video player sequentially for playback.
The maximum buffer size on user device is usually limited  (e.g., 20--40 Seconds).

\emph{\textbf{Notations:}}
Key notations in this part are listed below.

\vspace{1mm}
$\bullet$ $\tseg_{n}>0$:   segment length (in seconds) of user $n$'s video;

\vspace{1mm}
$\bullet$ $\R_n \eq \{R_n^1, R_n^2, ..., R_n^Z\}$ (with $0 < R_n^1 < R_n^2 < ... < R_n^Z$): the set of bitrates (in Mbps) available for user $n$, which depends on both the sever-side protocols and the user-side parameters such as device type.

\vspace{1mm}
$\bullet$ $ \Q_n > 0$: maximum buffer size (in seconds) of user~$n$.

%
%
%


\section{Problem Formulation}\label{sec:formulation}

In this section, we first characterize the users' behaviours in the cooperative streaming model, and then formulate the associated optimization problem.

Specifically, with the ABR streaming, each source video is downloaded \emph{segment by segment}. Namely, {each user starts to download a new segment (with a specific bitrate) only when completing the existing segment downloading.}
Hence, users operate in an \emph{asynchronous} manner, as they may complete  segment downloading at different times.
We refer to such an operation scheme as the \emph{segmented download operation}.

\subsection{Downloading Sequence}
With the segmented operation, each user $n$'s downloading operation can be characterized by a sequence:
\begin{equation}
\dset_n \eq \left\{\ds_{n[1]},\ \ds_{n[2]},\ ... ,\ \ds_{n[k]},\ ... \right\},
\end{equation}
with each element $\ds_{n[k]} $ denoting the information of the $k$-th downloaded segment, including
the segment owner $\du $,
bitrate level $\dz $,
bitrate $\dr = R_{\du}^{\dz}$,\footnote{{Here the bitrate $\dr = R_{\du}^{\dz}$ is redundant information, and mainly introduced for facilitating the later description.}}
download start time $\dts $, and end time $\dte $.
Namely, we can write $\ds_{n[k]} $ as
\begin{equation*}
\ds_{n[k]} = \left(\du ,\ \dz ,\ \dr,\ [\dts , \dte]  \right).
\end{equation*}
To distinguish the information of different segments, we will also write the information of segment $\ds_{n[k]}$ as $(\dunk, \dznk,  \drnk, \dtsnk, \dtenk)$ whenever needed.

It is easy to see that our cooperative streaming model generalizes the model without crowdsourcing, in which case we can simply restrict each user $n$ downloading only his own segment, i.e., $\dunk = n, \forall n, k$.

Next we provide the   constraints for a \emph{feasible} downloading sequence $\dset_n $ of user $n$.

(i) \emph{Timing Constraint:}
As users download segment by segment, we have the following timing constraint:
\begin{equation*}
\mathrm{C.1:}\quad
 \dtenk  \leq \dts_{n[k+1]} ,\quad  \forall k=1,..., |\dset_n  |. 
\end{equation*}
A strict inequality implies that user $n$ waits for some time before starting to download the next segment $ \ds_{n[k+1]} $, {for example, when all users' buffers are full.}


(ii) \emph{Capacity Constraint:} Each segment $\ds_{n[k]} =  (\du ,  \dz, \dr ,$ $  [\dts ,  \dte]   )$
consists of $\dr \cdot \tseg_{\du}    $ Mbits of video data, and is downloaded by user $n$ within time interval $\big[\dtsnk , \dtenk \big]$.
Hence, we have the following cellular link  capacity    constraint:
\begin{equation*}
\mathrm{C.2:}\quad
r \cdot  \tseg_{\du} \leq \int_{\dtsnk }^{\dtenk } \h_n(t) \dd t ,\quad  \forall k=1,...,|\dset_n|,
\end{equation*}
where $\h_n(t)$ is the real time cellular link capacity (in Mbps) of user $n$ at time $t$.

(iii) \emph{Encounter  Constraint:}
Each user can only download data for a nearby encountered user.
Hence, a segment with $\ds_{n[k]} =  (\du , \dz, \dr ,  [\dts ,  \dte]   )$, $n\neq \du$ is feasible only if users $n$ and $\du $ are encountered during $\big[\dtsnk ,\ \dtenk\big]$, i.e.,
\begin{equation*}
\mathrm{C.3:}\quad
e_{n,\du}(t) =1, \ t \in \left[\dtsnk ,\ \dtenk \right], \quad \forall k=1,...,|\dset_n| .
\end{equation*}

\subsection{Receiving Sequence}  \label{sec:form-receive}

Given the feasible downloading sequences of all users, i.e., $\dset_n, \forall n\in\N$, we can derive the segment receiving sequence of each user $m  $ as follows:\footnote{\revx{We assume the WiFi transmission time is zero.
One motivation for such an assumption is that the recent IEEE 802.11 standard family (e.g., 802.11n, ac, ad) has become increasingly powerful, and can support a data rate up to Gbit/s, which is much higher than those of the current cellular systems (such as 3G and 4G).
}}
\begin{equation}
\rset_m = \bigcup_{n\in\N,k\in\{1,\ldots,|\dset_n|\}:\dunk = m}  \left\{\ds_{n[k]}  \right\} .
\end{equation}
We assume that a proper download scheduling has been adopted, such that there is no repeated segments within $\rset_m$, and all segments in $\rset_m$ are sorted according to the playback order.  We denote the $k$-th segment in the reordered $\rset_m $ by $\rs_{m[k]}$. Then, we can write the receiving sequence of user $m$~as:
\begin{equation}
\rset_m \eq \left\{\rs_{m[1]},\ \rs_{m[2]},\ ... ,\ \rs_{m[k]}, \ ... \right\},
\end{equation}
with each element $\rs_{m[k]} = \left(\ru ,\ \rz,\ \rr ,\ [\rts , \rte]  \right) $ denoting the information of the $k$-th segment played by user $m$.
Similarly, we will write the information of $\rs_{m[k]}$ as $(\rumk ,  \rzmk ,  \rrmk ,  \rtsmk , \rtemk)$ whenever needed.

 It is easy to see that  $\rumk = m$  for all $\rs_{m[k]} \in \rset_m $.
To facilitate the later analysis, we further assume that $\rtemk \leq \rte_{m[k+1]}$, $\forall k=1,...,| \rset_m |$, that is, user $m$ receives the segments in $\rset_m$ sequentially.
 {Note that this can always be achieved by a proper schedule of downloading sequences with the full network information}. {For example, if $\rtemk > \rte_{m[k+1]} $, i.e., the $k$+$1$-th segment is received before the $k$-th segment, we can simply change their downloading orders.}


As mentioned previously, each received segment
is stored in a buffer at the user's device, and then is fetched to the video player sequentially for playback.
Let $\buf_{m[k]}$ denote the buffer level (in seconds) of user $m$ \emph{when receiving the $k$-th segment}, i.e., at the time $\rtemk$.
Then, we have the following \emph{\textbf{buffer update rule}} for user $m$:
\begin{equation}\label{eq:buffer-rule}
\buf_{m[k]} = \left[\buf_{m[k-1]} - \left(\rtemk - \rte_{m[k-1]} \right)\right]^+ + \tseg_m,
\end{equation}
where $[x]^+ = \max\{0,x\}$.
Here $\rtemk - \rte_{m[k-1]}$ is the time interval between   receiving of  $\rs_m[k-1] $ and $\rs_{m[k]} $, during which a period $\rtemk - \rte_{m[k-1]}$ of video is played back and removed from the buffer;
 $\tseg_m$ is the segment length (playback time) of the newly received segment $\rs_{m[k]}$.~~~~~~~~~

Since each user $m$'s buffer size is limited with $Q_m$ (seconds), we have the following \emph{buffer constraint}:
\begin{equation*}
\mathrm{C.4:}\quad
0 \leq \buf_{m[k]} \leq  Q_m,\quad \forall k=1,...,| \rset_m |.
\end{equation*}

\subsection{User Welfare}


The welfare of a user mainly consists of two parts:~a~\emph{utility} function capturing the user's QoE for video service, and a \emph{cost} function capturing the user's energy consumption for both video downloading and playing.

\emph{1) \textbf{Quality-of-Experience (QoE):}}
Users often desire  for a higher video quality without  frequent quality changes and freezes during playback.
Hence, a user's QoE mainly depends on the video quality,  quality fluctuation, and  rebuffering.
{Note that bitrate is a good measurement of video quality, and in general there is a distinct and monotonic relationship between bitrate and quality. Hence, we will define the QoE on bitrate for notational convenience.}

~~~~~~~~~~~~~~~~~

(i) \emph{Video Quality:}
A higher video  quality (bitrate) brings a higher value for users.
Let $g_n(r)$ denote the value that user $n$ achieves from  bitrate $r$ during one unit of playback time.
Then, the total value that user $n$ achieves from all received segments $\rset_n$ (each with a playback time of $\tseg_{\runk} = \tseg_n$) is:
\begin{equation} \label{eq:value}
\Va_n (\rset_n) \eq \sum_{k=1}^{|\rset_n|} g_n \left( \rrnk \right) \cdot \tseg_{n}.
\end{equation}
Obviously, $g_n(\cdot)$ is an increasing function  (as video quality monotonically increases with bitrate).
In our simulations, we adopt the following value function: $g_n(r) = \log(1 + \va_n \cdot r)$, where $\va_n >0 $ is a user-specific evaluation factor capturing user $n$'s desire  for a high quality video service.

(ii) \emph{Quality Fluctuation:}
The change of quality (bitrate) during playback decreases the user QoE, especially when the quality is degraded.
In this work, we assume that there is a value loss proportional to the bitrate decrease once the quality is degraded,
while there is  no value loss when the quality is upgraded.
Let $\ld_n > 0 $ denote the value loss of user $n$ for one unit (in Mbps) of bitrate decrease. Then, the total value loss of user $n$ induced by quality degradation is\footnote{Our model can be directly extended to the case with upgrade loss, by simply changing $[x]^+$ into the absolute operation $|x|$.}
 \begin{equation}\label{eq:loss-qd}
 \textstyle
\Lossdeg_n (\rset_n) \eq \sum\limits_{k=2}^{| \rset_n |} \ld_n \cdot \left[  \rr_{n[k-1]} -  \rrnk \right]^+,
\end{equation}
where $[x]^+ = \max\{0,x\}$. Here $\rr_{n[k-1]} >  \rrnk $ indicates that a quality degradation occurs between $\rs_n[k-1] $ and $\rs_{n[k]} $, with a bitrate decrease
 of  $\rr_{n[k-1]} -  \rrnk $.

(iii) \emph{Rebuffering:}
If a video buffer is exhausted before receiving a new segment, the video player   has to freeze the playback and rebuffer the video for a certain time. Such a freeze during playback is called \emph{rebuffering}.
The rebuffering during playback greatly affects the user QoE.
By the buffer update rule \eqref{eq:buffer-rule}, a rebuffering occurs when
\begin{equation*}
\buf_{n[k-1]} <  \rtenk - \rte_{n[k-1]},
\end{equation*}
with a detailed rebuffering time $\rtenk - \rte_{n[k-1]} - \buf_{n[k-1]}$.
Let $\lr_n >0 $ denote the value loss of user $n$ for one unit (second) of rebuffering time. Then, the total value loss of user $n$ induced by video rebuffering~is
 \begin{equation}\label{eq:loss-rebuf}
 \textstyle
\Lossrebuf_n (\rset_n) \eq \sum\limits_{k=2}^{| \rset_n |} \lr_n \cdot \left[ \rtenk -  \rte_{n[k-1]} - \buf_{n[k-1]} \right]^+.
\end{equation}

Based on the above, we can define the \emph{utility} of each user $n$ under a receiving sequence $\rset_n$   as follows:
\begin{equation}\label{eq:utility}
\Ut_n(\rset_n)\eq \Va_n (\rset_n)- \Lossdeg_n (\rset_n)- \Lossrebuf_n (\rset_n).
\end{equation}

\emph{2) \textbf{Energy Cost:}}
Users incur some energy cost in video streaming.
Such energy cost mainly includes the energy consumptions for downloading data via cellular links (and Internet) and exchanging data via   WiFi links.

(i) \emph{Energy Consumption for Video Downloading (via Celluar and Internet):}
When downloading data via the cellular link (and Internet), users' energy consumption depends on both the downloading time and the downloaded data volume \cite{energy}.
Let $\cct_n \geq 0$ denote the time-related energy consumption factor of user $n$ (i.e., for each unit  of downloading time), and $\ccv_n \geq 0$ denote the volume-related energy consumption factor of user $n$ (i.e., for each unit of downloaded data).
Then, the energy consumption of user $n$ for downloading video contents via cellular links and Internet is \cite{energy}:
\begin{equation}\label{eq:cost-c}
\textstyle
\Costc_n (\dset_n) \eq \sum\limits_{k=1}^{|\dset_n|}  \left(\cct_n \cdot (\dtenk - \dtsnk)
+ \ccv_n \cdot  \drnk \cdot \tseg_{\dunk}\right).
\end{equation}

(ii) \emph{Energy Consumption for Video Exchanging (via WiFi):}
When downloading a segment for others, the user needs to transmit the data to the segment owner via local WiFi link, the energy consumption of which also depends on the transmitting time and the transmitted data volume \cite{energy}.
Let $\cwt_n \geq 0$ and $\cwv_n \geq 0$ denote the time-related and volume-related energy consumption factors of user $n$ on the WiFi link, respectively.
The   energy consumption of user $n$  for video exchanging on WiFi link is \cite{energy}:
\begin{equation}\label{eq:cost-w}
\begin{aligned}
\textstyle
\Costw_n (\dset_n) \eq \sum\limits_{k=1}^{|\dset_n|}  & \left(\cwt_n \cdot 0
+ \cwv_n \cdot \drnk \cdot \tseg_{\dunk}\right)
\\
& \cdot  \textbf{1}( \dunk\neq n ),
\end{aligned}
\end{equation}
where $\textbf{1}( \dunk\neq n ) =1 $ if $\dunk\neq n$ (i.e., the segment $\ds_{n[k]}$ is downloaded for others), and $\textbf{1}( \dunk\neq n ) =0 $ otherwise.
Here we have assumed that the WiFi transmission time of a single segment is small and hence negligible.~~~~




Based on the above, we can derive the total \emph{energy consumption} of each user $n$ under a downloading sequence $\dset_n$ and receiving  sequence $\rset_n$ as follows:
\begin{equation}\label{eq:cost}
\Cost_n (\dset_n,\rset_n) \eq \Costc_n (\dset_n) + \Costw_n (\dset_n).
\end{equation}

\revx{Note that our proposed system can work with other energy models (e.g., those in \cite{energy-2}). In fact, the energy consumption modeling and energy saving are not the key objective of the proposed system. Instead, our key objective is to improve the QoE of users.}

\emph{3) \textbf{Welfare:}}
The {welfare} of each user $n$, denoted by $\Pay_n $, is defined as the difference between the utility (capturing the QoE of users) and the cost (capturing the energy consumption), i.e.,
\begin{equation}\label{eq:payoff}
\begin{aligned}
& \Pay_n ( \dset_n , \rset_n ) \eq \Ut_n (\rset_n)  - \Cost_n (\dset_n, \rset_n) . 
\end{aligned}
\end{equation}

The \emph{social welfare} is the aggregate welfare of all users:
\begin{equation}\label{eq:sw}
\textstyle
\SW (\dset_1, ..., \dset_N) \eq \sumN \Pay_n (\dset_n, \rset_n),
\end{equation}
where the receiving sequence $\rset_n $ of each user $n$ can be derived from the downloading sequences $\dset_n, n\in\N$.

%

\subsection{Problem Formulation}

Our purpose is to find the proper download scheduling to maximize the social welfare achieved in the proposed cooperative streaming model.

\revxx{First, in an ideal scenario with the complete network information,
we can formulate the following \emph{offline} social welfare maximization problem:
\begin{equation}\label{eq:swm}
\begin{aligned}
\max_{\{\dset_n, n\in\N\}} ~~& \SW (\dset_1, ..., \dset_N),
\\
\mbox{s.t.} ~~& \mathrm{C.1\sim C.4}.
\end{aligned}
\end{equation}
To solve this \emph{offline} optimization problem, we need to know the complete network information.
The solution of \eqref{eq:swm}, denoted by $\SWo$, provides the theoretical performance bound (in term of social welfare) of the proposed cooperative streaming system.
\revx{Note that \eqref{eq:swm} is an MILP (mixed integer linear programming) and  challenging to   solve.\footnote{\revx{This is because in each user's downloading sequence, we need to determine not only the order of segments, but also the download start time and end time for each segment. Even for the simplest case with a single user, it is still an NP-hard problem to optimally determine the download start time and end time of each segment.}}
Hence, we will derive a feasible upper-bound and a feasible lower-bound of $\SWo$ in Section \ref{sec:main1}.
}


Second, in a more general scenario without complete (future) network information, we need to design \emph{online}   scheduling algorithms, where the downloading operation of each user is performed in an online and distributed manner.
We will study such an online scheduling algorithm design and the associated performance evaluation in Section \ref{sec:main2}.}




\vspace{-1mm}

\section{Performance Bound Analysis}
\label{sec:main1}

In this section, we study the theoretical social welfare performance bound of the proposed cooperative system (i.e., the solution of the offline social welfare maximization problem \eqref{eq:swm}), which serves as a benchmark for the online scheduling solutions in Section~\ref{sec:main2}.

\revxx{However, directly solving  \eqref{eq:swm} is   challenging due to the following reasons.
First, users operate in an asynchronous manner.
Namely, different  users may start to download new segments at different times.  
Second,  \eqref{eq:swm} involves both discrete variables  (e.g., $\du$ and $\dz$) and  continuous  variables (e.g., $\dts$ and $\dte$), hence is a complicated mixed-integral  optimization problem.
Third, \eqref{eq:swm} involves the   integral operation ($\mathrm{C.2} $), which is even more challenging.
Hence,  we will focus on finding upper-bound and lower-bound for the desired performance bound of the cooperative streaming system.}



%

To achieve this, we propose a virtual \emph{time-slotted download operation} scheme, under which the problem can be formulated as an linear programming, hence can be solved by many classic methods.
We will show that the solution of \eqref{eq:swm} under the segmented operation scheme (i.e., the theoretical performance bound of the proposed cooperative streaming system) is bounded by the solutions under this virtual time-slotted system. {\textbf{It is important to note that this time-slotted operation scheme is only used for characterizing the theoretical performance bound, but not for the practical implementation.}}

\vspace{-1mm}

\subsection{Time-Slotted Download Operation}


To model the time-slotted operation scheme, we divide the whole time period $[0, T]$ into multiple time slots, each with the same   length.
For convenience, we normalize the length of each   slot to be one.
Hence, there is  a   set of $T$ time slots, denoted by $\T = \{1,2,...,T\}$, with the $\t$-th slot corresponding to   time interval $[\t-1,\ \t]$.

Under the time-slotted operation scheme, each video is downloaded \emph{slot by slot} in a {synchronized} manner, rather than segment by segment under the segmented operation.
Thus, in this case,  we can focus on the segments  that each user downloads in each time slot, instead of the segment downloading sequence.
Moreover, to guarantee the synchronous operation, we require that each segment must be completely downloaded within one time slot.
Namely, users cannot download a segment across multiple time slots.

\begin{figure}[t]
   \centering
  \includegraphics[height=1.15in]{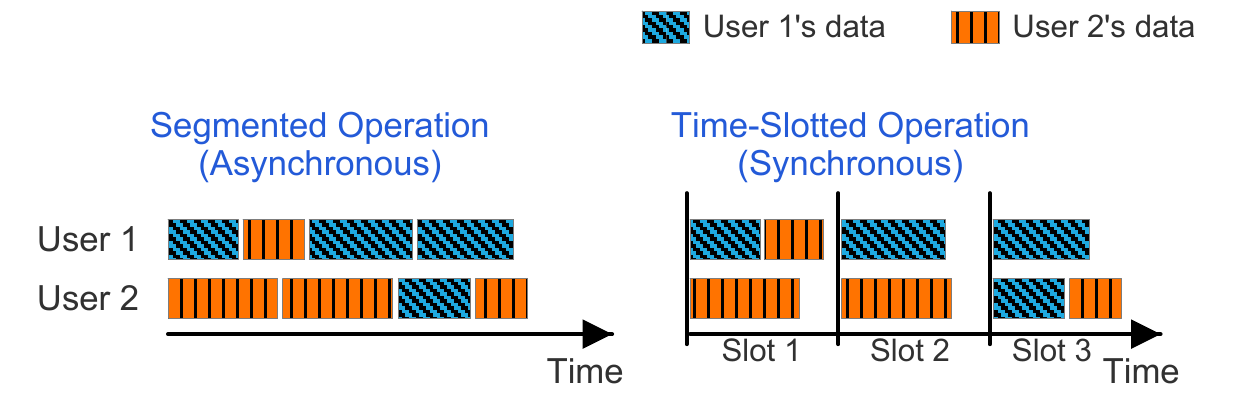}
 \caption{Segmented vs Time-Slotted Operation.}
  \label{fig:operation}
\vspace{-3mm}
\end{figure}

For clarity, we illustrate the difference (in download scheduling) between the segmented operation   and the time-slotted operation in Figure~\ref{fig:operation}, where blue blocks denote   user 1's data and orange blocks denote   user 2's data.
Under the segmented   operation  (left), users start to download data at different times, while under the time-slotted
  operation (right),
users are synchronized, and download data at the beginning of each time slot.~~~~~~~~~~~

\emph{1) \textbf{Downloading Vector:}}
With the time-slotted   operation,
the downloading operation of each user $n$ can be characterized by a downloading vector:
\begin{equation}
\kvec_n \eq \left\{  \k_{n,m}^z(\t), \ \ \forall  \t \in\T,  m\in\N, z\in\{1,...,Z\}  \right\},
\end{equation}
where each element $\k_{n,m}^z(\t)  $ is a non-negative integer,  denoting the total number of segments with bitrate level $z$ that user $n$ downloads for user $m$ in time slot $\t$.


Given the downloading vector $\kvec_n$,
we can derive the total data that user $n$ downloads  in each time slot $\t$:
\begin{equation} \textstyle
\xd_{n }(\t ) = \sumM  \x_{n,m}(\t) = \sumM    \sumZ \k_{n,m}^z(\t)  \cdot \tseg_m  \cdot  R_m^z ,
\end{equation}
where $\x_{n,m}(\t) \eq \sumZ \k_{n,m}^z(\t) \cdot \tseg_m \cdot  R_m^z $ is the amount of data   for user $m$ in slot $t$.
Then, we can define the link capacity constraint and encounter constraint for a feasible downloading vector $\kvec_n$:
\begin{equation*}
\begin{aligned}
\mathrm{\widetilde{C}.2:}\quad &
\xd_{n }(\t ) \leq  H_n({\t}),
\\
\mathrm{\widetilde{C}.3:}\quad &
e_{n,m}(t) = 1, t\in[\t-1, \t], \mbox{ if }	\x_{n,m}(\t) > 0,
\end{aligned}
\end{equation*}
where $H_n({\t}) =  \int_{\t-1 }^{\t } \h_n(t) \dd t$ is the total cellular link capacity of user $n$ in time slot $\t$.
Note that here we do not need to consider the timing constraint (C.1) as the    operation is already     slot by slot.

 \emph{2) \textbf{Receiving Vector:}}
Given feasible downloading vectors of all users, i.e.,
$\kvec_n,\forall n\in N$, we can derive the total playback time that user $m $ receives in each time slot $\t$:
\begin{equation} \textstyle
\yr_m(\t)  =  \sumN \y_{n,m}(\t) = \sumN    \sumZ \k_{n,m}^z(\t) \cdot \tseg_m  ,
\end{equation}
where $\y_{n,m}(\t) \eq \sumZ \k_{n,m}^z(\t) \cdot \tseg_m$ is the total playback time that user $m$ receives from user  $n$ in slot $\t$.

Let $\buf_m(\t)$ denote the buffer level (in seconds) of user $m$ \emph{at the end of time slot $\t$}.
Then, we have the following \emph{\textbf{buffer update rule}} for user $m$:
\begin{equation}\label{eq:buffer-rule-slot}
\buf_m(\t)  = \left[\buf_m(\t-1) - 1 \right]^+ + \yr_m(\t),
\end{equation}
where $[x]^+ = \max\{0,x\}$.
Here one time unit of video is played back during time slot $\t$, and $\yr_m(\t)$ is the playback time of the newly received segments in slot $\t$.

Similarly, we have the following buffer constraint:
\begin{equation*}
\mathrm{\widetilde{C}.4:}\quad
0 \leq \buf_m(\t) \leq  Q_m,\quad \forall \t=1,...,T.
\end{equation*}

 \emph{3) \textbf{User Welfare:}}
Now we define the user welfare and social welfare under the time-slotted operation.

(i) \emph{Video Quality:}
Similar as \eqref{eq:value}, the   value that user $n$ achieves from all received segments is:
\begin{equation} \label{eq:value-slot}
\textstyle
\xVa_n  \eq  \sumTa \sumM \sumZ \k_{m,n}^z(\t)  \cdot \tseg_n  \cdot g_n(R_n^z) .
\end{equation}

(ii) \emph{Quality Fluctuation:}
Without loss of generality, we assume that all the received segments of each user $n$ in each time slot $\t$ are sorted in ascending order of bitrate.\scfootnote{If not, we can simply change the orders of related segments.}
Hence, quality degradation only occurs between two successive time slots, while never occurs within a time slot.
Let ${r}_n^{\textsc{h}}(\t)$ and $r_n^{\textsc{l}}(\t)$ denote the highest bitrate and lowest bitrate that user $n$ receives in slot $\t$.
 Then, similar as \eqref{eq:loss-qd}, the   value loss of user $n$ due to quality degradation~is:
 \begin{equation}\label{eq:loss-qd-slot}
 \textstyle
\xLossdeg_n \eq \sum\limits_{\t=2}^{T} \ld_n \cdot \left[ r_n^{\textsc{h}}(\t-1) -  r_n^{\textsc{l}}(\t) \right]^+ .
\end{equation}

(iii) \emph{Rebuffering:}
By the buffer update rule in \eqref{eq:buffer-rule-slot}, a rebuffering occurs in time slot $\t$ when
$$
\buf_m(\t-1) < 1 ,
$$
with a rebuffering time $1 - \buf_m(\t-1)$.
Then, similar as \eqref{eq:loss-rebuf}, the value loss of user $n$ induced by rebuffering~is
 \begin{equation}\label{eq:loss-rebuf-slot}
\textstyle \xLossrebuf_n \eq \sum \limits_{\t=2}^{T} \lr_n \cdot \left[ 1  -\buf_m(\t-1) \right]^+.
\end{equation}

(iv) \emph{Energy Consumption for Video Downloading (via Cellular and Interent):}
Similar as \eqref{eq:cost-c}, the energy consumption  of user $n$ for downloading video is
\begin{equation}\label{eq:cost-c-slot}
\textstyle \xCostc_n \eq \sumTa  \left(\cct_n \cdot \frac{\xd_{n }(\t )}{H_n({\t})}
+ \ccv_n \cdot  \xd_{n }(\t )  \right),
\end{equation}
where $\frac{\xd_{n }(\t )}{H_n({\t})}$ is the actual downloading time in   slot $\t$.

(v) \emph{Energy Consumption for Video Exchanging (via WiFi):}
Similar as \eqref{eq:cost-w}, the energy consumption of user $n$ for exchanging video on local WiFi links is
\begin{equation}\label{eq:cost-w-slot}
\textstyle\xCostw_n \eq \sumTa  \sum\limits_{m=1, m\neq n}^N
\left(\cwt_n \cdot 0
+ \cwv_n \cdot \x_{n,m}(\t)   \right).
\end{equation}


Based on the above,  the welfare of each user $n$  is
\begin{equation}\label{eq:payoff-slot}
\begin{aligned}
 \xPay_n (\kvec_1, ..., \kvec_N) \eq \xVa_n - \xLossdeg_n - \xLossrebuf_n
  - \xCostc_n - \xCostw_n .
\end{aligned}
 \end{equation}


 \emph{4) \textbf{Problem Formulation under Time-Slotted Operation:}}
 Now we can define the social welfare maximization problem under the time-slotted download operation:
\begin{equation}\label{eq:swm-slot}
\begin{aligned}
\max_{\{\kvec_n, n\in\N\}} ~~ & \textstyle \xSW   \eq \sumN \xPay_n(\kvec_1, ..., \kvec_N) ,
\\
\mbox{s.t.} ~~& \mathrm{\widetilde{C}.2\sim \widetilde{C}.4}.
\end{aligned}
\end{equation}
Similar to \eqref{eq:swm}, this is an \emph{offline} optimization problem and requires the complete network information.
Moreover, \eqref{eq:swm-slot} is an integer programming,
and can be   solved by many classic methods.
Hence, we skip the detailed derivations.
For notation convenience, we denote the solution of \eqref{eq:swm-slot} by $\SWx$.

%

\subsection{Performance Bound}


Now we characterize the theoretical performance  bound $\SWo$ of the proposed cooperative streaming system (under the segmented operation) by using the solution $\SWx$ of \eqref{eq:swm-slot} under the virtual time-slotted operation.

For convenience, we denote $\btseg \eq (\tseg_1, ..., \tseg_N)$ as the   vector consisting of all users' segment lengths, and denote  $\SWo_{(\btseg)}$ and  $\SWx_{(\btseg)}$
 as the    solutions of \eqref{eq:swm} and \eqref{eq:swm-slot} under  $\btseg$, respectively.
 {We refer to a   vector $\btseg$ as an \emph{integer multiple} of another vector $\btseg'$, if each element $\tseg_n$ in $\btseg$ is an integer multiple of the corresponding element $\tseg_n'$ in $\btseg'$}.
 For example, $\btseg = (1,...,N)$ is an
 integer multiple of $\btseg' = (0.5,...,N/2)$.

\begin{proposition}
If $\btseg$ is an  {integer multiple} of   $\btseg'$, then
$$
\SWo_{(\btseg)}  \leq  \SWo_{(\btseg')}, \mbox{~~~and~~~}
\SWx_{(\btseg)}  \leq  \SWx_{(\btseg')}.
$$
\end{proposition}

This proposition can be proved by showing that in both schemes, any downloading operation under $ \btseg $ can be equivalently achieved under $\btseg'$.

\begin{proposition}
If $\btseg \rightarrow \mathbf{0}$ (i.e., $\tseg_n \rightarrow 0, \forall n\in\N$), then
$$
\SWo_{(\btseg)} = \SWx_{(\btseg)}.
$$
\end{proposition}

This proposition can be proved by showing that with infinitely small segment lengths $\btseg \rightarrow \mathbf{0}$, any downloading operation under the time-slotted operation scheme can be equivalently achieved under the segmented operation scheme, and vise versa.

\begin{proposition}
If $\btseg \succeq \mathbf{0}$ is a finite vector (i.e., each element $\tseg_n \geq 0$ is a finite number), then
$$
\SWo_{(\btseg)} \geq  \SWx_{(\btseg)}.
$$
\end{proposition}

This proposition can be proved by showing that with finite segment lengths $\btseg \succeq \mathbf{0}$, any downloading operation under the time-slotted operation scheme  can be equivalently achieved under the segmented operation scheme, but \emph{not}   vise versa.

Based on the above, we have the following theorem.
\begin{theorem}
Given a segment length $\btseg $, the theoretical performance upperbound $\SWo_{(\btseg)}$ is bounded by:
$$
\SWx_{(\btseg)} \leq   \SWo_{(\btseg)} \leq  \SWx_{(\btseg' \rightarrow \mathbf{0})}.
$$
\end{theorem}




Intuitively, this theorem states that with any  $\btseg $, the theoretical performance bound $\SWo_{(\btseg)}$ of our proposed cooperative streaming system is
(a) lower-bounded by $\SWx_{(\btseg)}$ (i.e., the optimal performance of the virtual time-slotted system with the same segment length vector $\btseg $), and   (b) upper-bounded by $\SWx_{(\btseg' \rightarrow \mathbf{0})}$ (i.e., the optimal performance of the virtual time-slotted system with infinitely small segment lengths $\btseg' \rightarrow \mathbf{0} $).
Therefore, the performance of the virtual time-slotted system under different $\btseg $ characterizes the theoretical performance region  of our proposed cooperative streaming system.


\vspace{-1mm}

\section{Online Scheduling Algorithms}
\label{sec:main2}

In  the  previous section, we have analyzed the theoretical performance bound of the  cooperative streaming system, which is achievable in an ideal scenario with complete  network information.
In practice, however, network changes randomly over time, and hence it is difficult  to  obtain the future and global network information.

In this section, we study the practical scenario where the future and global network information is not available.
We propose an online scheduling algorithm based on the Lyapunov optimization framework \cite{lya}, which relies only on the current local  network information and the scheduling history, while not on any future or global network information.

\vspace{-1mm}

\subsection{Online vs Offline}

We first discuss the key difference between online scheduling and offline scheduling.
\revxx{In the offline scheduling, the segment downloading sequences of all users at all time are determined in advance, through, for example, the offline social welfare maximization problem \eqref{eq:swm}, which requires the complete network information.
In the online scheduling, however, each user makes the download scheduling decision (regarding the next segment to be downloaded) in real time, e.g., at the time when he completes a previous segment downloading.}~~~~~~~~~~~~~~~~~~~~

In our proposed cooperative streaming system, such a real time downloading decision mainly includes two problems: \emph{whose segment to be downloaded, and at which bitrate level?}
The decision may depend on different criteria such as
the real time user buffer levels (e.g., in \cite{b1}),
the channel bandwidth or throughput predictions (e.g., in \cite{b6}),
and other specific objective functions (e.g.,
Lyapunov drift-plus-penalty described below).~~~~

\vspace{-1mm}

\subsection{Lyapunov-Based Online Scheduling}


Lyapunov optimization \cite{lya} is a widely used technique for
solving stochastic optimization problems with time average
constraints.
In our model, an implicit time average constraint is that
the average segment arriving rate should be same as the video playback rate in term of segment.\scfootnote{For example, for a video with 2-second segment, the playback rate in term of segment is 0.5 (segments per second).}
\revh{If the video playback rate is smaller, then the downloaded segments will be frequently dropped due to the limited buffer size;
if the video playback rate is larger, then the rebuffering will frequently happen.
Both cases are not desirable in this system.}
To this end, we introduce the Lyapunov optimization technique to optimize the downloading scheduling  in an online manner.

Suppose that a user $n$ completes a segment downloading at time $t$, and needs to make the downloading decision regarding the next segment to be downloaded.
We denote such a decision  by $(u, z)$, where $u\in \N$ is the owner of the segment to be downloaded, and $z\in \{1,...,Z\}$ is the bitrate level of the segment to be downloaded.
Obviously, a feasible decision $(u,z)$ of user $n$ at time $t$  satisfies the following user encounter constraint: $e_{n,u}(t) = 1$.

For analytical convenience, we further denote $\buf_m (t) $ as the  buffer level  of each user $m$ at time $t$,
and denote $r_m $ as the bitrate of user $m$'s \emph{last received}  segment.  This information   captures the current network state and historical scheduling information  that can be observed.

\emph{1) \textbf{Objective Function:}}
Given a feasible decision $(u, z)$ of user $n$, the data volume to be downloaded is $R_u^z  \cdot \tseg_u$ (Mbit), and the \emph{estimated} downloading time is $\gamma_{u,z} \eq \frac{R_u^z  \cdot \tseg_u}{h_n(t)}$.\scfootnote{Here we use the current channel capacity $h_n(t)$ to approximate the capacity in a period of future time.
Note that the actual downloading time may be different from $\gamma_{u,z} $ due to the channel stochastics.}
\revv{The total energy consumption of user $n$ (\emph{for this particular   downloading operation}) and user $u$ (for playing the downloaded segment)  is:}
\begin{equation*}
 \begin{aligned}
&  \Cost_n(u,z)  =    \Costc_n + \Costw_n .
\end{aligned}
\end{equation*}

The utility of receiver $u$ \emph{on this particular segment} is
\begin{equation*}
\begin{aligned}
 \Ut_u(u,z) &  = \Va_u- \Lossdeg_u- \Lossrebuf_u .
\end{aligned}
\end{equation*}

The utility of other user $m \neq u $ due to this operation~is
\begin{equation*}
 \textstyle  \Ut_m(u,z)  = - \Lossrebuf_m=
-    \lr_m \cdot  \left[  \gamma_{u,z}  -  \buf_m (t) \right ]^+ , ~~
\end{equation*}
which only includes the potential rebuffering loss.

Therefore, the total   welfare generated under $(u, z)$ is
\begin{equation}\label{eq:obj} \textstyle
\Pay(u, z) \eq  \sumM \Ut_m(u,z)  - \Cost_n(u,z).
\end{equation}


\emph{2) \textbf{Lyapunov Drift:}}
Following the Lyapunov framework, we   define a modified Lyapunov function:
 \begin{equation} \textstyle
 J(t) \eq \frac{1}{2}   \sumM \left[\Q_m -   \buf_m (t) \right] ^2.
\end{equation}
The \emph{Lyapunov drift} is  the change of Lyapunov function (from one decision-making time to the next), i.e.,
 \begin{equation} \label{eq:drift}
\begin{aligned}
\Delta(t) & \eq  J(t +  \gamma_{u,z}) - J(t),
\\
& \textstyle = \frac{1}{2} \sumM \left( \left[\Q_m -   \buf_m (t + \gamma_{u,z}) \right]^2 -   \left[\Q_m -   \buf_m (t) \right]^2 \right),
\end{aligned}
\end{equation}
where $\buf_m (t + \gamma_{u,z})$ is the estimated buffer level of user $m$ at time $t + \gamma_{u,z}$
(i.e., the next decision-making time of user $n$).
For the   receiver  $u$,  the estimated buffer level is:
$$
\buf_u (t + \gamma_{u,z}) = \min\{\Q_u,\ [\buf_u (t) - \gamma_{u,z}]^+ +  \tseg_u \}.
$$
For other user $m \neq u$, the estimated buffer level is:
$$
\buf_m (t + \gamma_{u,z}) = [\buf_m (t) - \gamma_{u,z}]^+  .
$$

\emph{3) \textbf{Online Scheduling Algorithm:}}
By the Lyapunov optimization theorem, to stabilize the system while optimizing the objective, we can use such a scheduling policy that greedily minimizes \emph{drift-plus-penalty}:
 \begin{equation} \label{eq:drift-penalty}
\Phi(t) \eq  \Delta(t) - \lambda \cdot  \Pay(u, z),
\end{equation}
where  the negative welfare $(-\Pay(u, z))$ is viewed
as the penalty incurred at time $t$, and  $\lambda \geq 0$ is a control parameter.
It is important to note that the buffer levels (appearing in $\Delta(t)$) serve as regulation factors, such that the user
with a larger idle buffer can be more likely to be scheduled (hence
reducing the possibility of rebuffering).
This term is different from the rebuffering loss in (\ref{eq:loss-rebuf}), which is the actually realized loss when a rebuffering event
actually happens.

\begin{algorithm}[t]
\label{algo:Lya}
\caption{Lyapunov-based Online Scheduling}
\DontPrintSemicolon
\While{at each decision-making time $t$}
{
\If{$\buf_n(t) + \tseg_n > \Q_n , \forall n\in \N$}
{
/* \emph{no buffer can afford one more segment}  */
\\
Wait for $T_w =  \min_{n\in\N} (\buf_n(t) + \tseg_n - \Q_n)$  seconds;
}
\Else
{
Download a segment of bitrate level $z^*$  for user $u^*$:
\\
$ ~~~~~
(u^*, z^*) = \arg \min_{u , z  } \  \Phi(t) \eq  \Delta(t) - \lambda \cdot  \Pay(u, z)
$
}
}
\end{algorithm}

Based on the above analysis, we now design an
on-line algorithm that aims at minimizing the drift-plus-penalty
\eqref{eq:drift-penalty} in each decision-making time.
We present the detailed algorithm in Algorithm \ref{algo:Lya}.
Note that a user may decide \emph{not} to
download any segment at a decision-making time, when, for example, all buffers are full and cannot afford one more segment.
In this case, the user will wait for a certain time and then trigger decision-making event again.
Hence, \emph{a decision-making time can be either the time that a user   completes a segment downloading}  or \emph{the time that a user  is triggered by the waiting timer}.~~~~~~~~~~~~~


Note that the online scheduling in Algorithm \ref{algo:Lya} works in a distributed manner, as each user makes the decision independently.
To coordinate the downloading decisions of different users and to avoid the redundant downloading of the same segment, nearby users need  to exchange the context information (e.g., buffer length, segment size, encode bitrate, and url). To illustrate this, we construct a real demo system on Raspberry PI.
Please refer to our online technical report \cite{report} for more details.

\emph{4) \textbf{Performance Analysis:}}
Now we analyze the performance of Algorithm \ref{algo:Lya}.
Let $t_{[k]}$ denote the $k$-th decision-making time (counting  all users), and let $\Pay_{[k]}$ denote the associated welfare achieved in the $k$-th download operation.
Then, the social welfare generated by Algorithm \ref{algo:Lya} during the whole time $[0,T]$  can be computed~by:
$$
\textstyle
\SWi_{(\btseg)} = \sum\limits_{t_{[k]} \leq T} \Pay_{[k]}.
$$
By the Lyapunov optimization theorem  (Theorem 4.2 in \cite{lya}), we obtain the following gap for $\SWi_{(\btseg)}$ and $\SWo_{(\btseg)} $, i.e., the theoretical performance upperbound.
\begin{theorem} \label{theorem:2}
$$
\lim_{T \rightarrow \infty}
 \
\mathrm{E} \left[ \SWi_{(\btseg)} \right]
 \geq
\mathrm{E} \left[ \SWo_{(\btseg)}  \right] - \frac{B}{\lambda},
$$
where $\mathrm{E}[.]$ is expectation, and $B$ is a positive constant.
\end{theorem}

Theorem \ref{theorem:2} shows that Algorithm \ref{algo:Lya} converges to the theoretical performance bound  $\SWo_{(\btseg)}$  asymptotically, with a controllable approximation error bound $O(\frac{1}{\lambda})$.

However, this theorem does not directly help us calculate the \emph{actual gap} between $\SWi_{(\btseg)}$ and $\SWo_{(\btseg)} $ in a particular experimental scenario, as $T$ is finite  in practice.
To this end, we propose another approach based on Theorem 1 for the practical calculation of the actual gap, i.e.,
$$
\big| \SWo_{(\btseg)} - \SWi_{(\btseg)} \big|
\leq
\big| \SWx_{(\btseg' \rightarrow \mathbf{0})} -\SWi_{(\btseg)} \big|.
$$
Note that $ \SWx_{(\btseg' \rightarrow \mathbf{0})}$ is the solution of the integer programming problem (\ref{eq:swm-slot}), and can be easily computed in a practical experiment after collecting the complete network information.
In our experiments, the average gap between $\SWi_{(\btseg)}$ and $\SWx_{(\btseg' \rightarrow \mathbf{0})}$ is smaller than $3\%$.

%
%
%

%


\section{Experiments and Performance}\label{sec:simulation}

%
%

%

\begin{figure*}[t]
 \vspace{-5mm}
 \centering
  \includegraphics[width=.23\textwidth]{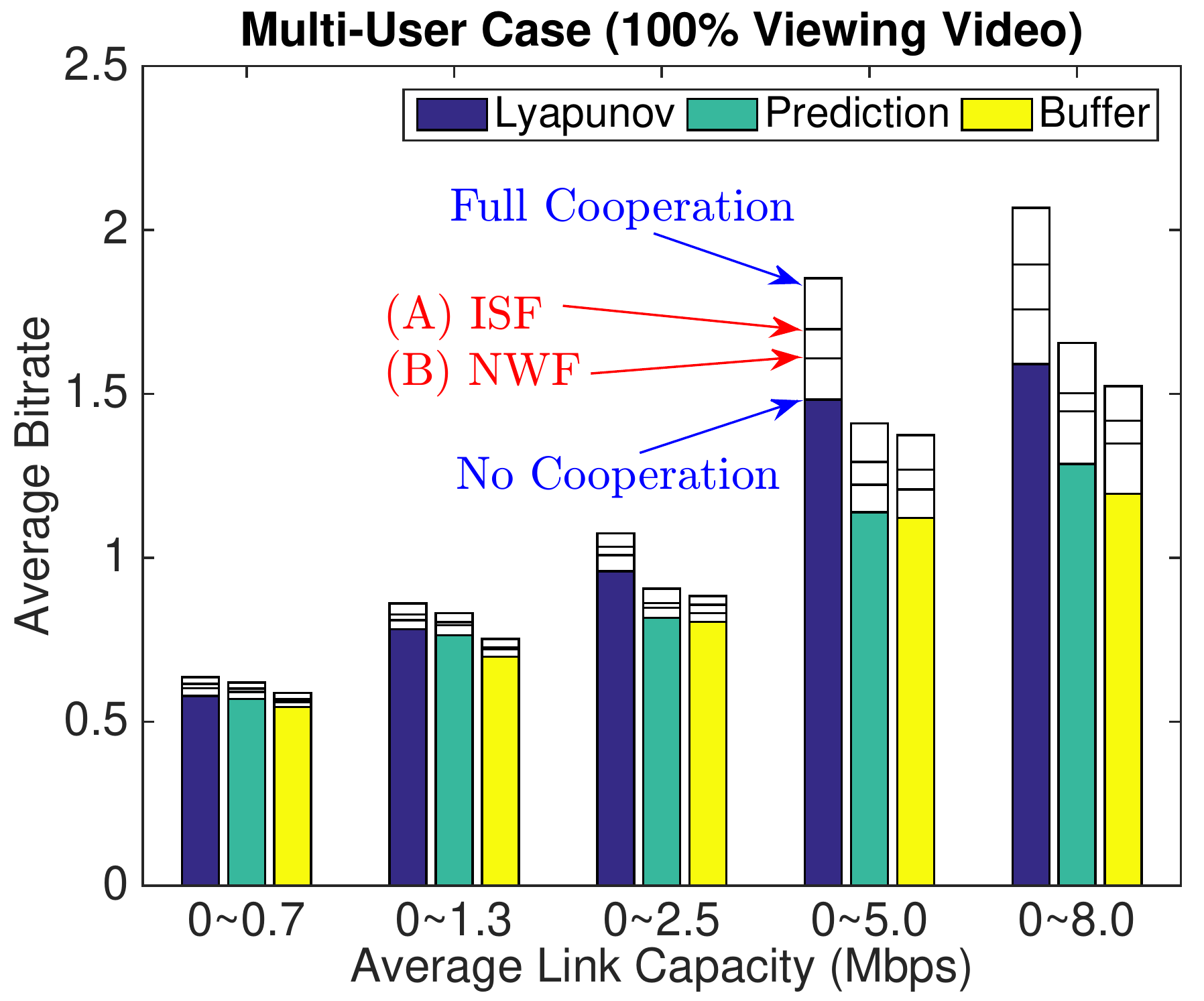}
  ~
  \includegraphics[width=.23\textwidth]{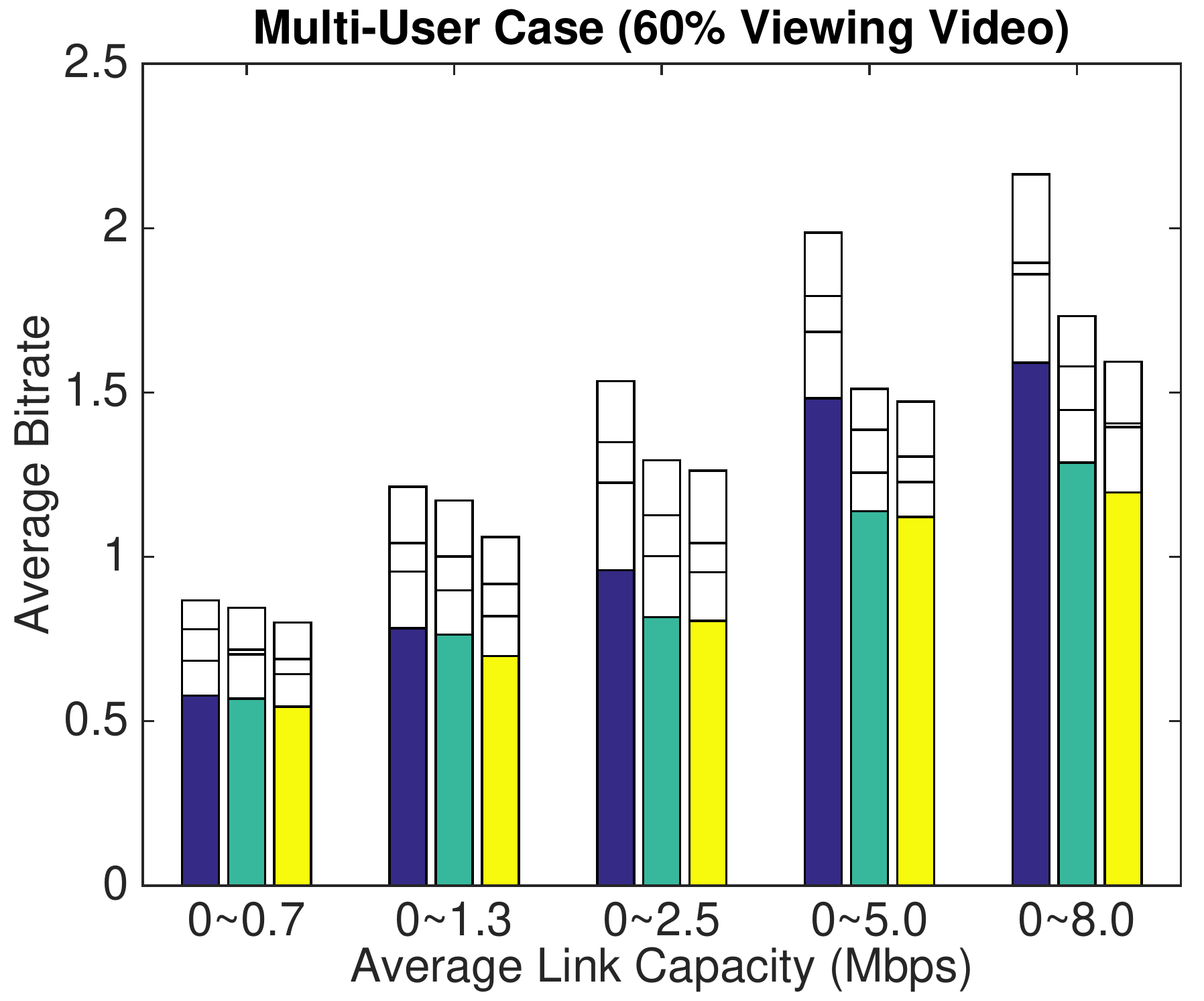}
  ~
  \includegraphics[width=.23\textwidth]{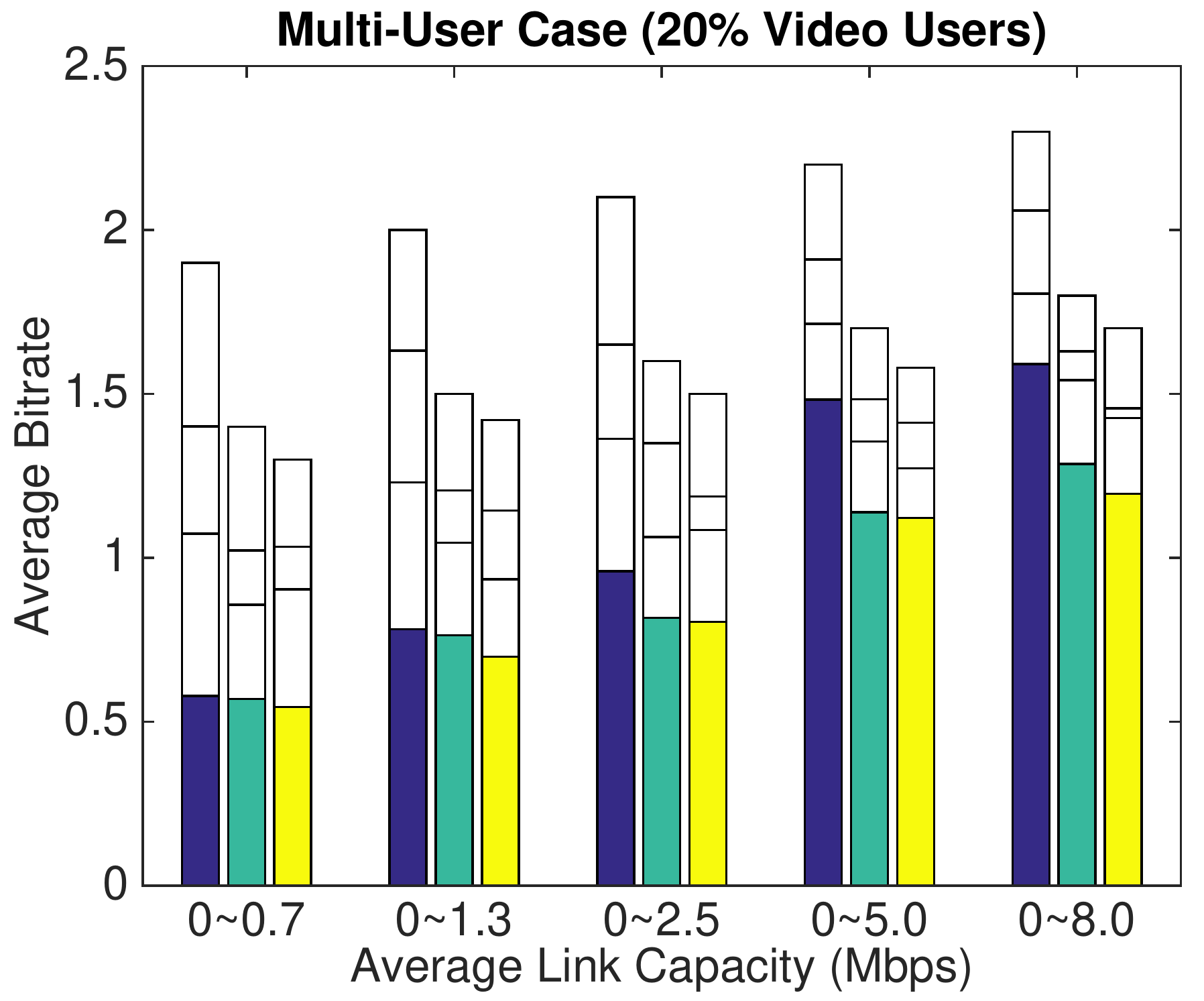}
  ~
  \includegraphics[width=.23\textwidth]{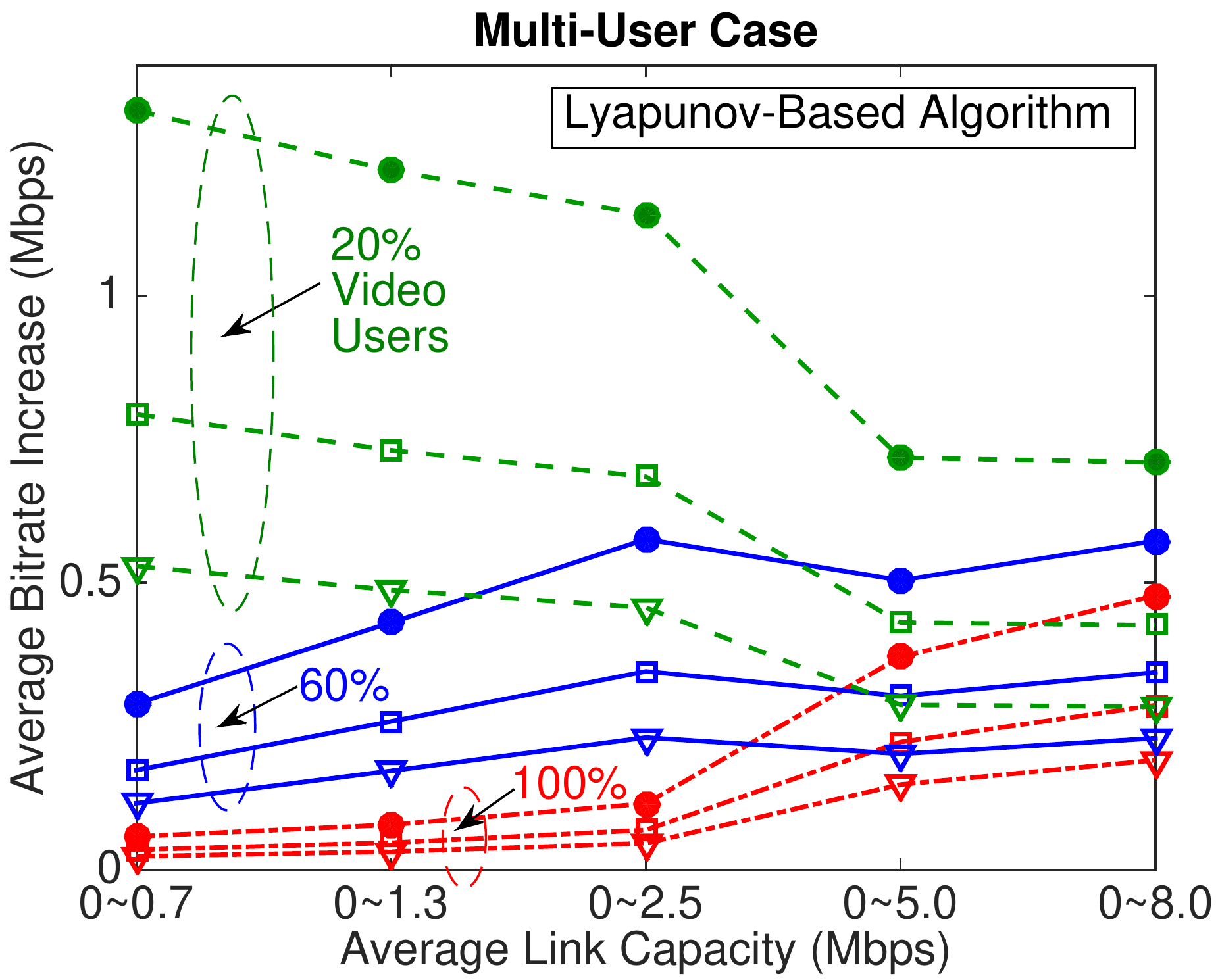}
  \vspace{-1mm}
\\
~~~~~~~~
~~~~~
(a)
~~~~~~
~~~~~~~~
~~~~~~~~
~~~~~~~~
(b)
~~~~~
~~~~~~~~
~~~~~~~~
~~~~~~~~
(c)
~~~~~~~~
~~~~~
~~~~~~~~
~~~~~~~~
(d)
~~~~~~~~
~~
\vspace{-2mm}
  \caption{(a) Average Bitrate with 100\% Video Users, (b) Average Bitrate with 60\% Video Users, (c) Average Bitrate with 20\% Video Users, (d) Average Bitrate Increase under the Lyapunonv-based Algorithm.}
  \label{fig:multiple-bitrate}
\vspace{-3mm}
\end{figure*}

\subsection{Experiment Setting}

 {\emph{1) \textbf{Datasets:}}
To evaluate the realistic performance of the proposed cooperative streaming system, we conduct experiments based on real data traces from two datasets: ISF \cite{data1} and NWF \cite{data2}.\footnote{ISF is provided by a non-profit organization   ``Ile Sans Fil'' in Canada, and is open source (available at CRAWDAD   \cite{data1}).
NWF is obtained from a wireless service provider ``NextWiFi'' in China \cite{data2}.}
Both datasets record the user access sessions at a set of WiFi hotspots in different countries during a long period of time (3 years for ISF and 5 months for NWF),
representing two different (hotspot-based) mobility scenarios: users encounter more frequently in NWF, while the duration of each encounter is larger in ISF.}

To simulate the video watching behaviours of mobile users and the real cellular link throughputs for video streaming, we use the video viewing session logs obtained from BestTV \cite{data3}, one of the largest OTT (Over The Top) video service providers in China.
There are 5 different bitrate levels (for mobile users) in this dataset: $\{0.2, 0.4, 0.7, 1.3, 2.3\}$Mbps, corresponding to the lowest to the highest video resolutions, respectively.
Based on the segment length, bitrate, and downloading time, we can calculate the \emph{measured} end-to-end link throughput for each
segment downloading.
We use this measured throughput to approximate the cellular link capacity in our experiments.
Moreover, the energy consumption factors
are chosen according to the real measurement given in  \cite{smartphone-energy}.

%
%

{\emph{2) \textbf{Existing Online Algorithms:}}
To evaluate the performance of our proposed Lyapunov-based online algorithm, we also perform simulations using the following two typical existing online algorithms: Buffer-based algorithm  \cite{b1} and Channel Prediction-based algorithm   \cite{b6}.
Specifically,   buffer-based algorithm \cite{b1} introduces a linear mapping between buffer and bitrate, and selects the next segment bitrate based on the current buffer level: \emph{a higher buffer level is mapped to a higher bitrate}.
Channel prediction-based algorithm \cite{b6} proposes a channel prediction method, and selects the next segment bitrate based on the predicted channel capacity: \emph{the highest bitrate that can be supported by the predicted channel capacity}.\footnote{Note that both algorithms in \cite{b1}  and \cite{b6}  were designed for the single-user scenario, and considered only the bitrate adaptation.
In the multi-user scenario, we need to consider both   bitrate adaptation and   segment owner selection (i.e., whose segment to be downloaded) as discussed in Section \ref{sec:main2}.
To this end, we introduce the following segment owner selection policy for these two algorithms in the multi-user scenario:
\emph{Each user $n$ will choose to download the next segment for  another user $u \neq n$, if and only if (i) $\buf_n \geq \delta_{\textsc{th}} \cdot \Q_n$, (ii) $ \buf_n -\buf_u \geq \Delta_{\textsc{th}} $, and (iii) $ \buf_u = \min_{m\in\N} \buf_m $.}
Intuitively, user $n$ will choose to help the user with the lowest buffer level, if his own buffer level is higher than a ratio threshold $\delta_{\textsc{th}}$ and meanwhile is higher than the lowest buffer level by  a threshold $\Delta_{\textsc{th}}$. In our experiments, we will try different values of $\delta_{\textsc{th}} $ and $\Delta_{\textsc{th}}  $, and choose the best ones.}}

%

\subsection{Multiple-User Case}

%
%
%
%
%
 Now we perform experiments for the multi-user scenario, where some users play  videos (called video users), while others remain idle and can potentially help the encountered video users.\footnote{We also construct experiments for the single-user scenario (i.e., non-cooperative scenario) to illustrate the performance gap of our proposed Lyapunov-based online algorithm to the theoretical performance bound as well as to compare the bitrate adaptation performance of our proposed algorithm with the existing online algorithms. The detailed results are provided in our online technical report \cite{report}.}
For simplicity, we assume that all video users play the high-resolution videos (bitrate $2.3$Mbps).
The total video length is 500 seconds,  the segment length is 2 seconds, and the maximum  buffer length at the user's device is 40 seconds.
We use these multi-user experiments to illustrate both the cooperation gain of the proposed cooperative streaming system and the performance gain of the proposed algorithm.

In the following experiments, we consider a total of 50 users and randomly choose a subset of users as video users.
We consider different network conditions, characterized by the \emph{range} of the average link capacity.
For example, a bad network condition corresponds to a range $[0, 0.7]$Mbps, under which each user will be randomly assigned by a real data trace with an average link capacity smaller than $0.7$Mbps.

\emph{1) \textbf{Average Bitrate:}}
Figure \ref{fig:multiple-bitrate} shows the average bitrates with different percentages of video users under different network conditions.
For each video user percentage and network condition, we perform experiments with the three algorithms under ISF and NWF mobility traces, corresponding to different encountering scenarios (hence different cooperation probabilities).
To fully characterize the cooperation gain, we also run the algorithms under two benchmark encountering scenarios: (i) a full cooperation scenario, where all users are always encountered with each other, and (ii) a non-cooperative scenario, where none of users are encountered.

Sugfigures (a) to (c) show the average bitrates with $100\%$, $60\%$, and $20\%$ video users, respectively.
As illustrated in (a), the solid bar denotes the average bitrate under the non-cooperative scenario, and the hollow bar denotes the average bitrate under the full cooperation, in which the first (higher) line denotes the average bitrate under ISF (with a higher encountering probability) and the second (lower) line denotes the average bitrate under NWF (with a lower encountering probability).
Subfigure (d) shows the average bitrate \emph{increase} (i.e., the cooperation gain) using our proposed Lyapunov-based algorithm, comparing with the achieved bitrate under the non-cooperative scenario.
The dash, solid, and dash-dot lines denote the results with $20\%$, $60\%$, and $100\%$ video users, respectively. The marks ``circle'', ``square'', and ``triangle'' denote full cooperation, ISF, and NWF, respectively.

From subfigure (a), we can see that when the percentage of (high-resolution) video users is very high (e.g., $100$\%), the  increase  of bitrate is very small under a low link capacity range (e.g., lower than $2.5$Mbps), as in this case all users are lack of capacity, hence nobody can help other users significantly.
Under a high link capacity range (e.g., $[0, 5]$Mbps and $[0, 8]$Mbps), the  increase  of bitrate becomes significant, as some users may have redundant capacities, hence can help others.
From subfigures (b) and (c), we can see that when the percentage of video users is low (e.g., $60\%$ or $20\%$), the  bitrate increase is significant under all network conditions, mainly due to the contributions of the idle users.

Subfigure (d) summarizes the increase of bitrate under our proposed algorithm.
We can see that with $100\%$ video users, the increase of bitrate continuously increases with the link capacity, as a larger capacity gives the video users more opportunities to obtain redundant capacity and help others.
With $20\%$ video users, however, the increase of bitrate continuously decreases with the link capacity, as a very small capacity already leads to a considerably high bitrate (due to the contributions of a large population of idle users), hence the increase of bitrate is more significant under a small capacity (as the benchmark bitrate is smaller).
With $60\%$ video users, the increase of bitrate first increases with the link capacity (due to a similar reason in the $100\%$ case), and then decreases with the link capacity (due to a similar reason in the $20\%$ case).
The maximum bitrate increase ratio under the full cooperation scenario can be up to $50\%\sim230\%$  with $20\%$ video users,
$35\%\sim60\%$ with $60\%$ video users, and $4\%\sim40\%$  with $100\%$ video users.
Moreover, the bitrate increase under the real data traces is bounded by the above maximum ratio, and actually depends on the encountering probability. In our experiments, the bitrate increases under ISF and NWF can reach around $60\%$ and $40\%$ of the maximum bitrate increase, respectively.~~~~

\begin{figure*}[t]
 \vspace{-5mm}
  \centering
  \includegraphics[width=.23\textwidth]{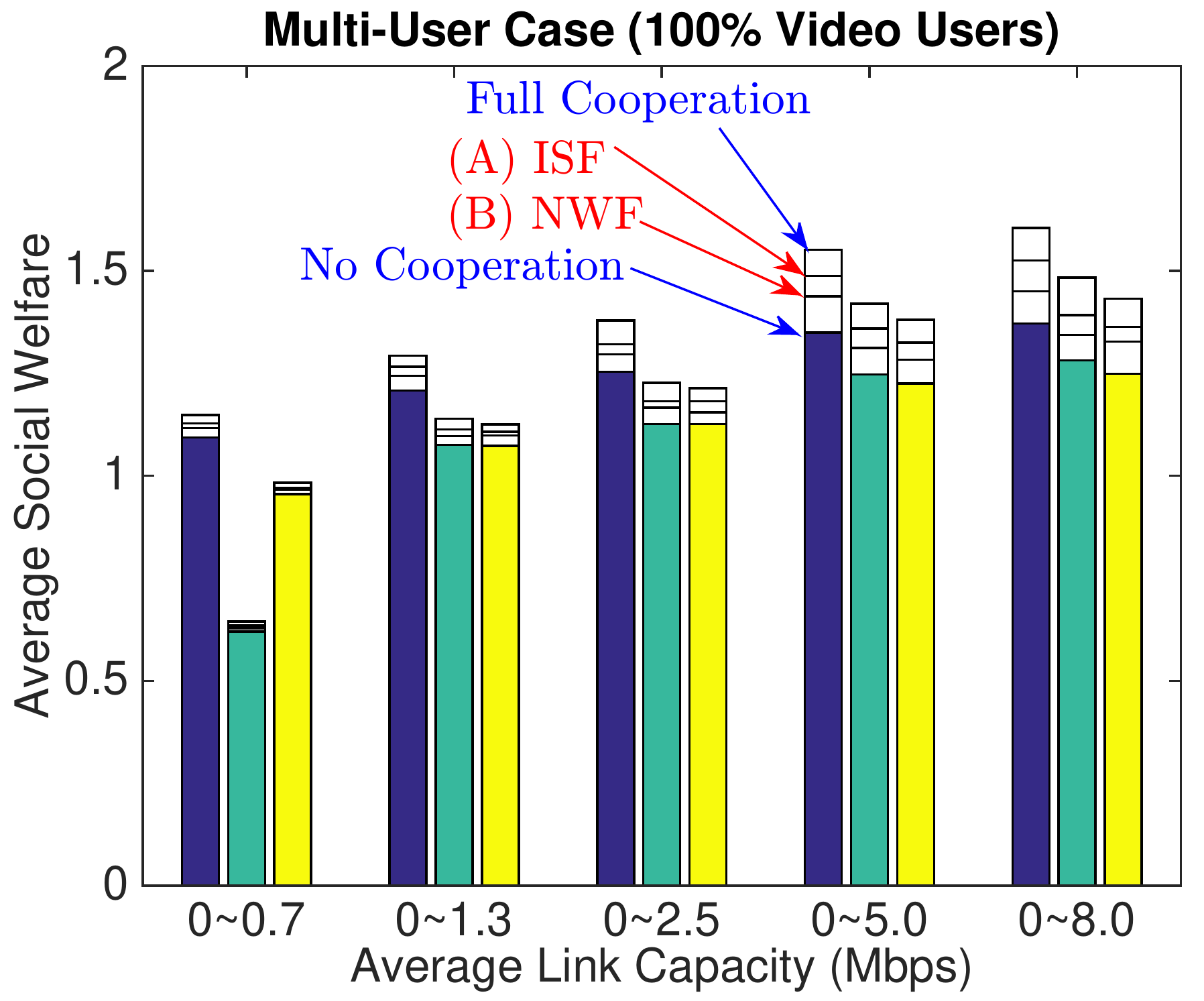}
  ~
  \includegraphics[width=.23\textwidth]{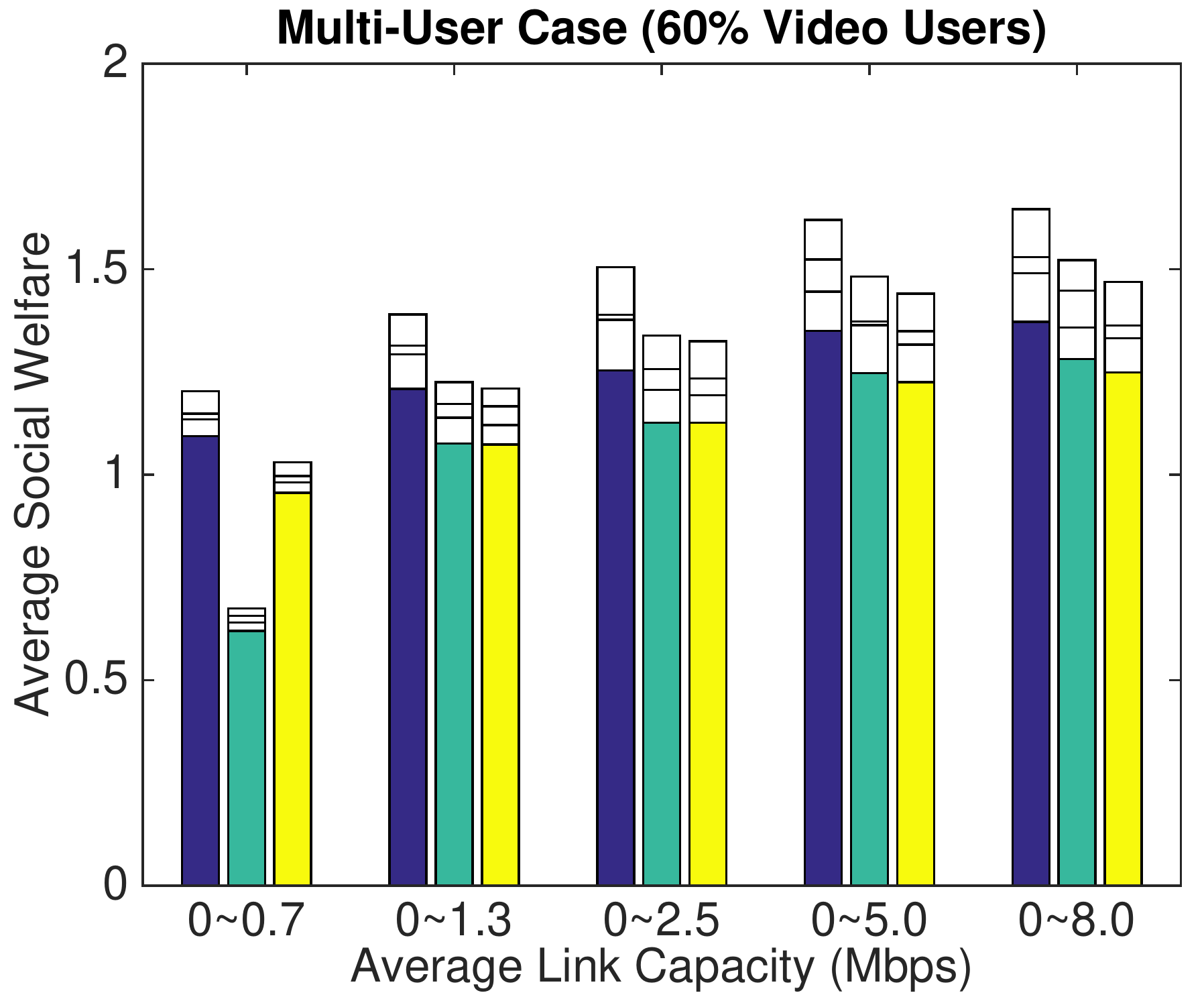}
  ~
  \includegraphics[width=.23\textwidth]{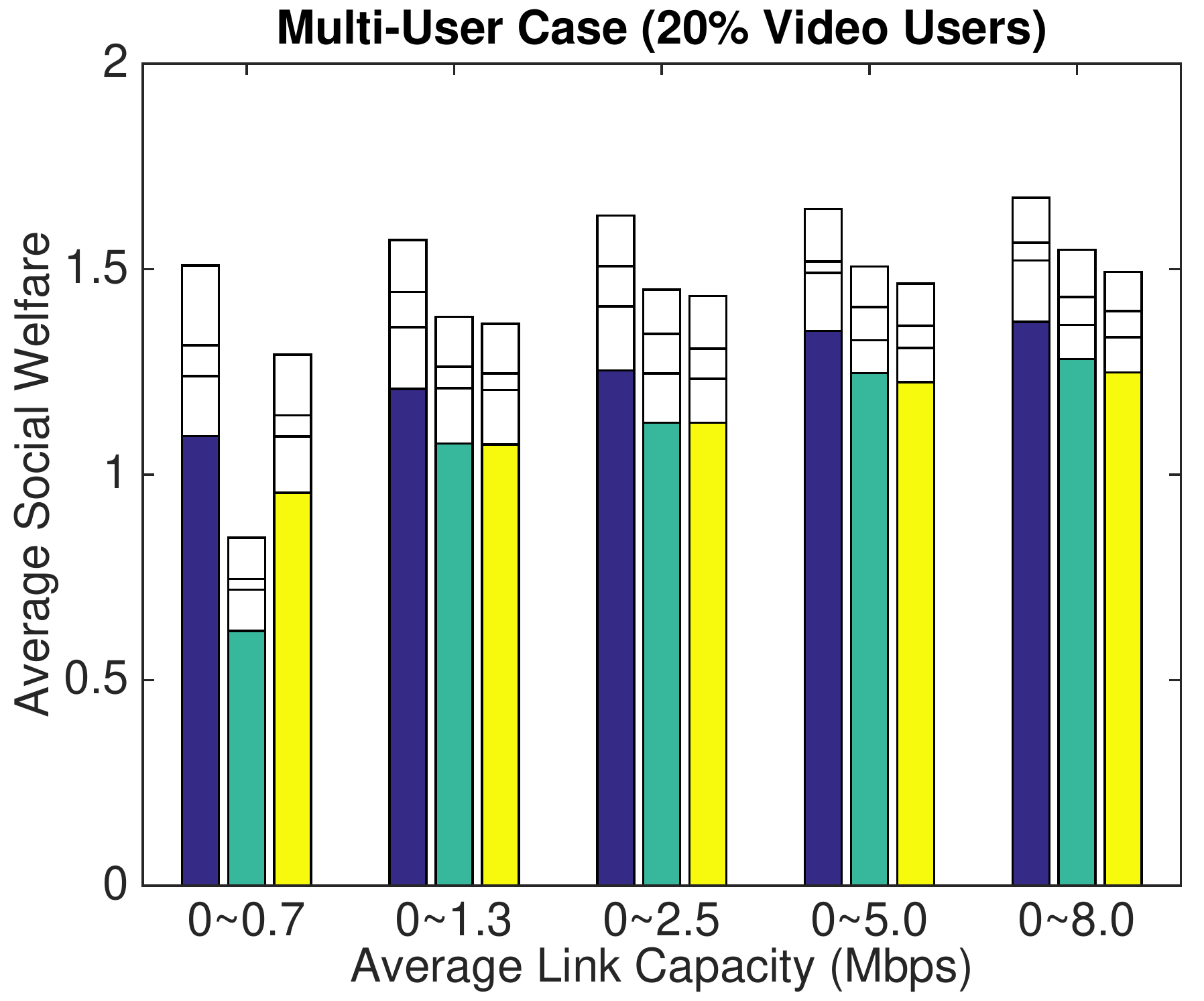}
  ~
  \includegraphics[width=.23\textwidth]{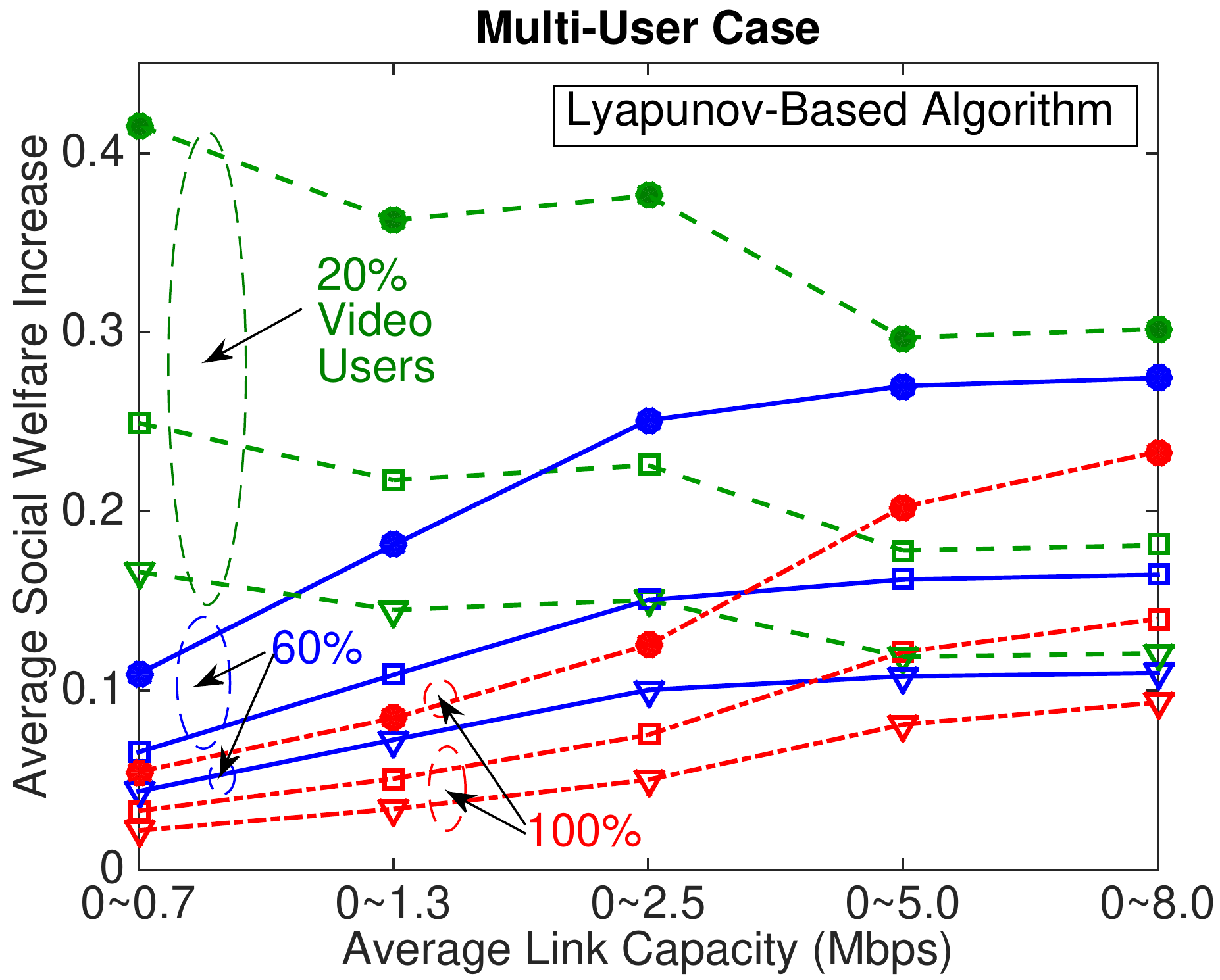}
  \vspace{-1mm}
\\
~~~~~~~~
~~~~~
(a)
~~~~~~
~~~~~~~~
~~~~~~~~
~~~~~~~~
(b)
~~~~~
~~~~~~~~
~~~~~~~~
~~~~~~~~
(c)
~~~~~~~~
~~~~~
~~~~~~~~
~~~~~~~~
(d)
~~~~~~~~
~~
\vspace{-2mm}
  \caption{(a)   Social Welfare with 100\% Video Users, (b)   Social Welfare with 60\% Video Users, (c)   Social Welfare with 20\% Video Users, (d)  Average Social Welfare Increase  under the Lyapunonv-based Algorithm.  }
  \label{fig:multiple-sw}
\vspace{-3mm}
\end{figure*}

\emph{2) \textbf{Social Welfare:}}
Figure \ref{fig:multiple-sw} shows the average social welfares and welfare gains with different percentages of video users under different network conditions.
The key informations and observations regarding the social welfare are similar as those regarding the average bitrate in Figure \ref{fig:multiple-bitrate}, hence we skip the detailed discussions and only present the results regarding the cooperation gain.
Specifically, using our proposed algorithm, the maximum social welfare increase ratio (under the full cooperation scenario) can be up to $20\%\sim40\%$ with $20\%$ video users,
$10\%\sim20\%$ with $60\%$ video users,
and
$5\%\sim15\%$ with $100\%$ video users.
The social welfare increase under ISF and NWF can reach $60\%$ and $40\%$ of the maximum welfare increase, respectively.~~~~~~~~~~~~~~~


\emph{3) \textbf{Algorithm Comparison:}}
From Figure \ref{fig:multiple-bitrate} (a) to (c) and
Figure \ref{fig:multiple-sw} (a) to (c), we can also evaluate the performance difference between our proposed algorithm and the   algorithms in \cite{b1} and \cite{b6} in the multi-user scenario.
By comparing the difference between solid bars (for the non-cooperative scenario) and the difference between hollow bars (for the multi-user cooperative scenario), we can find that the performance difference (between our algorithm and the algorithms in \cite{b1, b6}) become more significant in the cooperative  scenario, especially when the video user percentage is small.
Such a performance difference  is mainly due to the non-optimal segment owner selection in \cite{b1, b6}. In our algorithm, however, the segment owner selection and the bitrate adaptation are optimised jointly.

\section{Conclusion}
\label{sec:con}

In this work, we proposed a   multi-user cooperative video streaming framework for video streaming over wireless networks.
We analyzed the theoretical performance bound of the proposed cooperative streaming system, and designed the online  streaming algorithm for the practical implementation.
We conducted extensive experiments with real data traces, and illustrated both the cooperation gain of the cooperative streaming  system and the performance gain of the proposed online streaming  algorithm.
Adaptive bitrate streaming is a new technology trend of mobile video streaming, and the research on multi-user cooperative video streaming is becoming increasingly important.
This paper developed a unified cooperative framework, for both theoretical analysis and practical implementation.
There are several interesting future research directions in this area.
An important one is to consider the users' strategic behaviours and the associated incentive issues in the cooperative streaming.

\section{Acknowledgments}

This work is supported by the National Natural Science Foundation of China (Grant No. 61771162, 61472204, and 61521002) and the General Research Funds (Project Number CUHK 14219016) established under the University Grant Committee of the Hong Kong Special Administrative Region, China.
This work is also supported by the Beijing Key Lab of Networked Multimedia (Z161100005016051).
Jianwei Huang is the corresponding author.


\begin{IEEEbiography}[{\includegraphics[width=1in,clip,keepaspectratio]{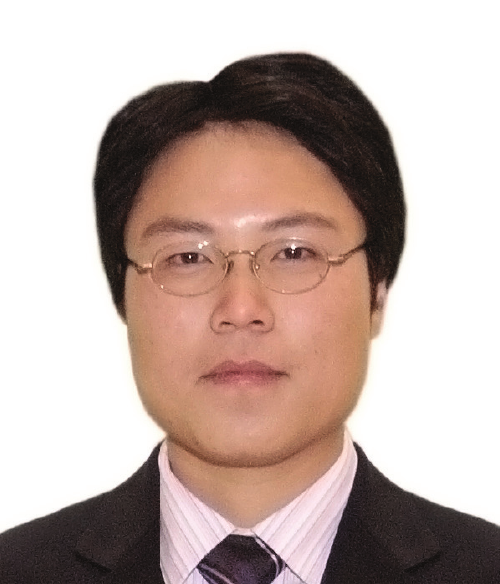}}]{Lin Gao}
(S'08-M'10-SM'16) is an Associate Professor
with the School of Electronic and Information
Engineering, Harbin Institute of Technology,
Shenzhen, China.
He received the Ph.D. degree
in Electronic Engineering from Shanghai Jiao
Tong University in 2010.
His main research interests
are in the area of network economics and
games, with applications in wireless communications
and networking.
He received the IEEE ComSoc Asia-Pacific Outstanding Young Researcher Award in 2016.
\end{IEEEbiography}

\begin{IEEEbiography}[{\includegraphics[width=1in,clip,keepaspectratio]{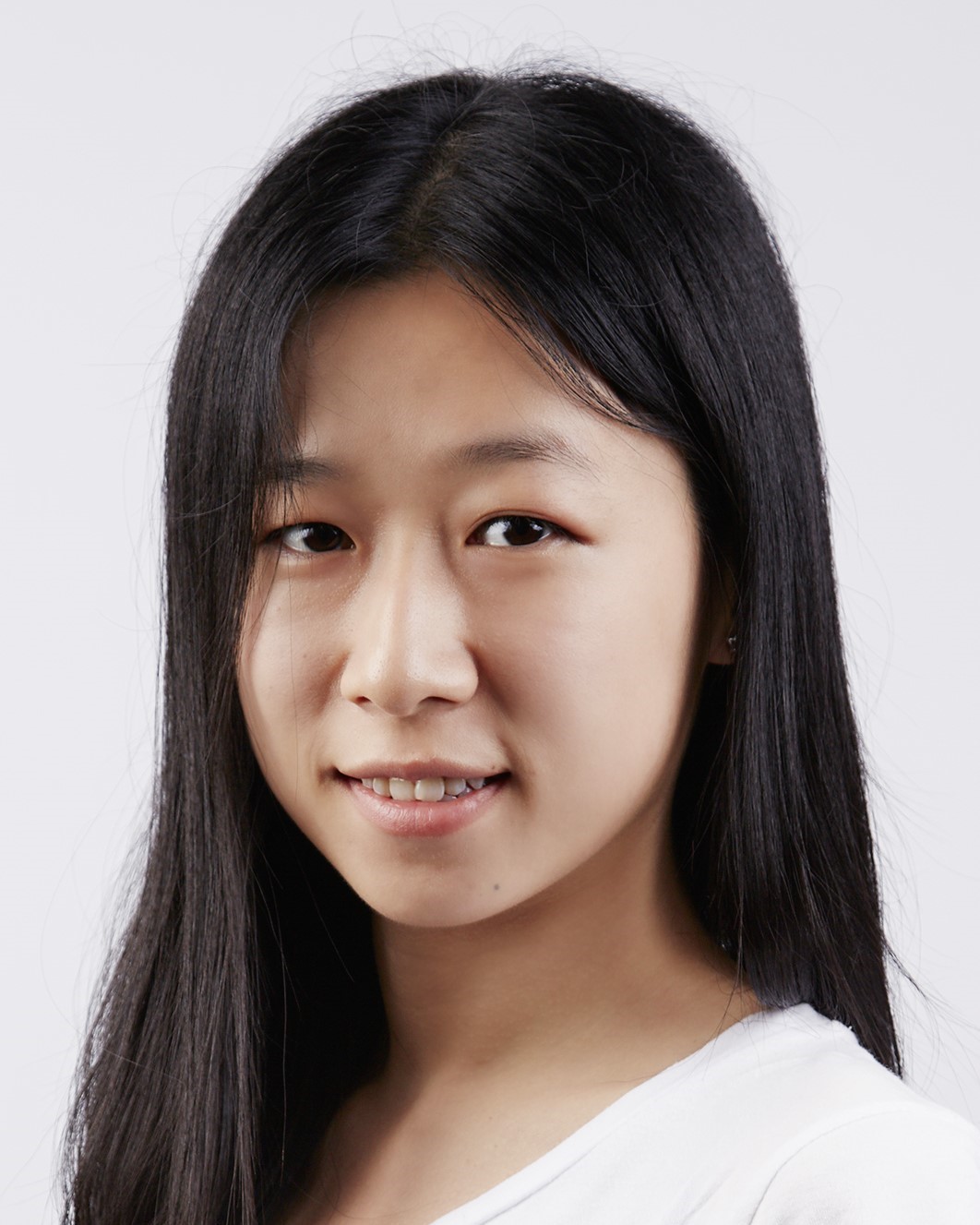}}]{Ming Tang}
(S'16)  is currently pursuing a Ph.D. degree at the Department of Information Engineering, The Chinese University of Hong Kong (CUHK). Her research interests include wireless communications and network economics, with particular emphasis on user-provided networks, mobile video streaming, and fog computing.
\end{IEEEbiography}

\begin{IEEEbiography}[{\includegraphics[width=1in,clip,keepaspectratio]{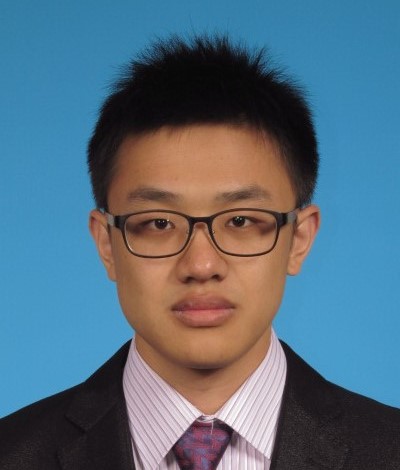}}]{Haitian Pang}
(S'16) received his BE degree in Department of Automation in 2014 from Tsinghua University, Beijing, China. He is currently a PhD candidate in Computer Science in Tsinghua University. His research areas include network game modeling, cellular-WiFi networking, video streaming system design, and mobile networking optimizations.
\end{IEEEbiography}

\begin{IEEEbiography}[{\includegraphics[width=1in,clip,keepaspectratio]{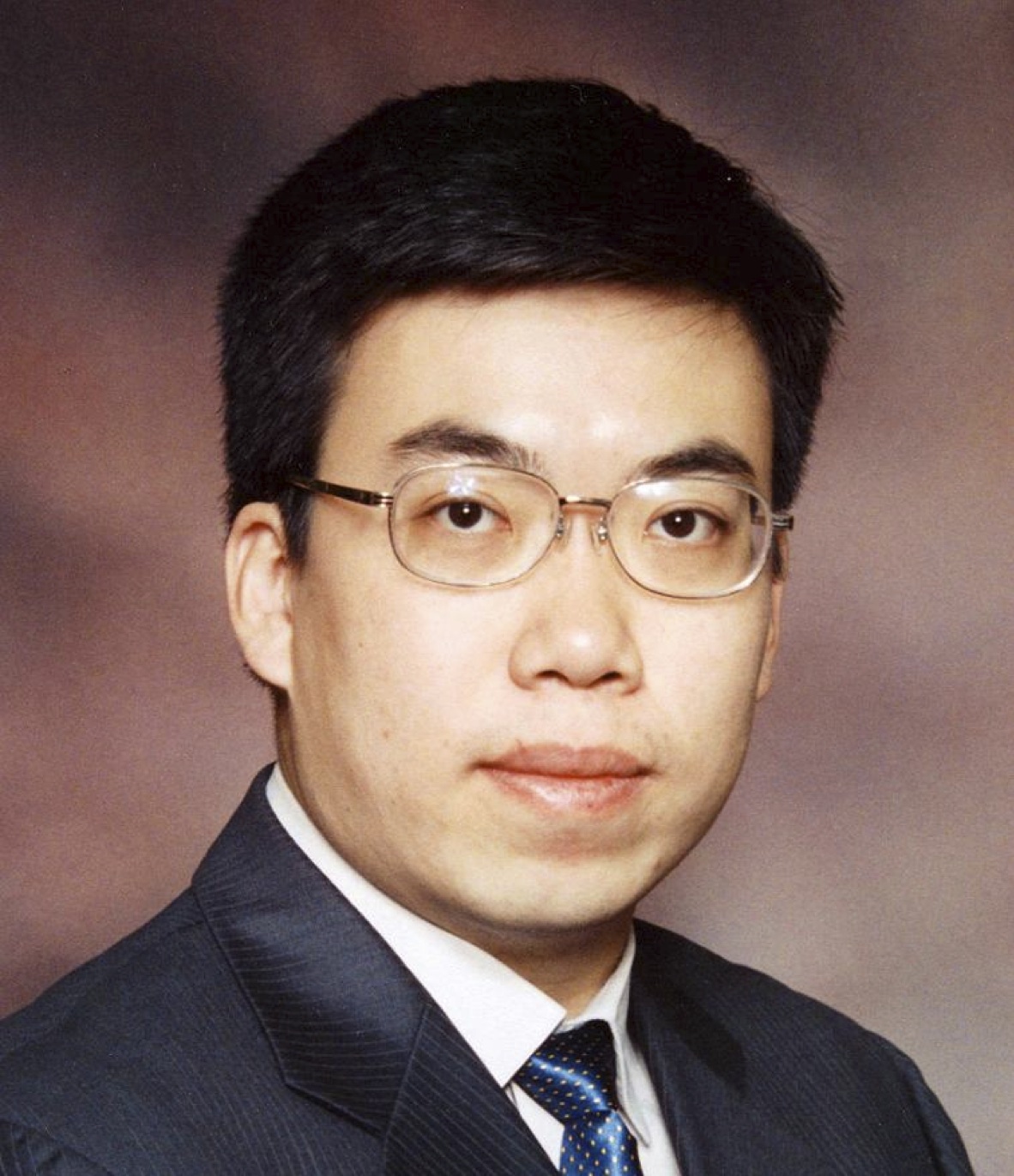}}]{Jianwei Huang}
(F'16) is a Professor in the Department of Information Engineering at The Chinese University of Hong Kong. He is the co-author of 9 Best Paper Awards, including IEEE Marconi Prize Paper Award in Wireless Communications 2011. He has co-authored six books, including the textbook on ``Wireless Network Pricing''. He has served as the Chair of IEEE TCCN and MMTC. He is an IEEE ComSoc Distinguished Lecturer and a Thomson Reuters Highly Cited Researcher.
\end{IEEEbiography}

\begin{IEEEbiography}[{\includegraphics[width=1in,clip,keepaspectratio]{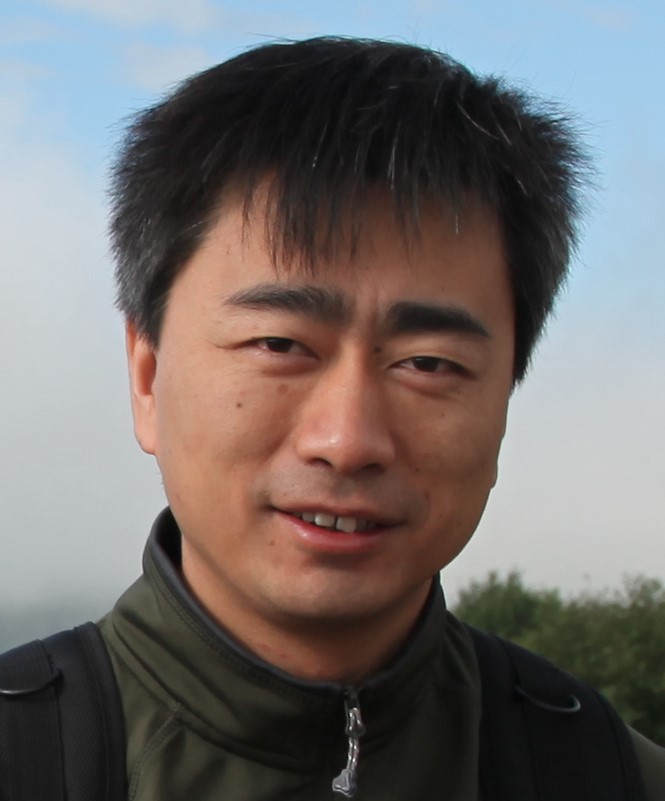}}]{Lifeng Sun}
was born in 1972. He received the Ph.D. degree in System Engineer from the National University of Defense Technology, Changsha, in 2000. Currently, he is a professor at Tsinghua University. His research interests include video streaming, video coding, video analysis and multimedia cloud computing. He is a member of IEEE and ACM. He received Best Paper Award in IEEE Transactions on Circuits and Systems for Video Technology in 2010, Best Paper Award at ACM Multimedia 2012, and Best Student Paper Award at MMM 2015 and IEEE BigMM 2017.
\end{IEEEbiography}

\newpage

\appendix

\subsection*{A.1 Implementation on Demo System}

To illustrate the implementation of the proposed cooperative streaming system, we construct a demo system on Raspberry PI Model B+ (with the Wheezy-Raspbian operating system).\footnote{For more details about Respberry PI, please refer to: http://www.raspberrypi.org.}
In the demo system, Raspberry PIs act as the mobile devices in the practical system, which are equipped with monitors (for video playing),
LTE USB modems (for LTE connections), and WLAN
adapters (for WiFi connections).\footnote{\revaa{Note that we implement the demo system on Raspberry PI to simulate its implementation and operation on real smartphones. The key reasons for such a simulation are following. First, Raspberry PI is more  user-friendly in programming on almost all
functionalities, while some functionalities of smartphones (both Android and IOS) are not easily programmable. Second, such a simulation on Raspberry PI is able to capture the key features of a real system on smartphones (when it is developed).}}
The devices can dynamically join and leave the cooperative group, and there is no need for centralized control. After joining the cooperative group,
the mobile devices download video segments via LTE and
forward video segments as well as control messages to other devices
(if needed) through WiFi.

\begin{figure}[t]
  \centering
 \includegraphics[height=3.5in]{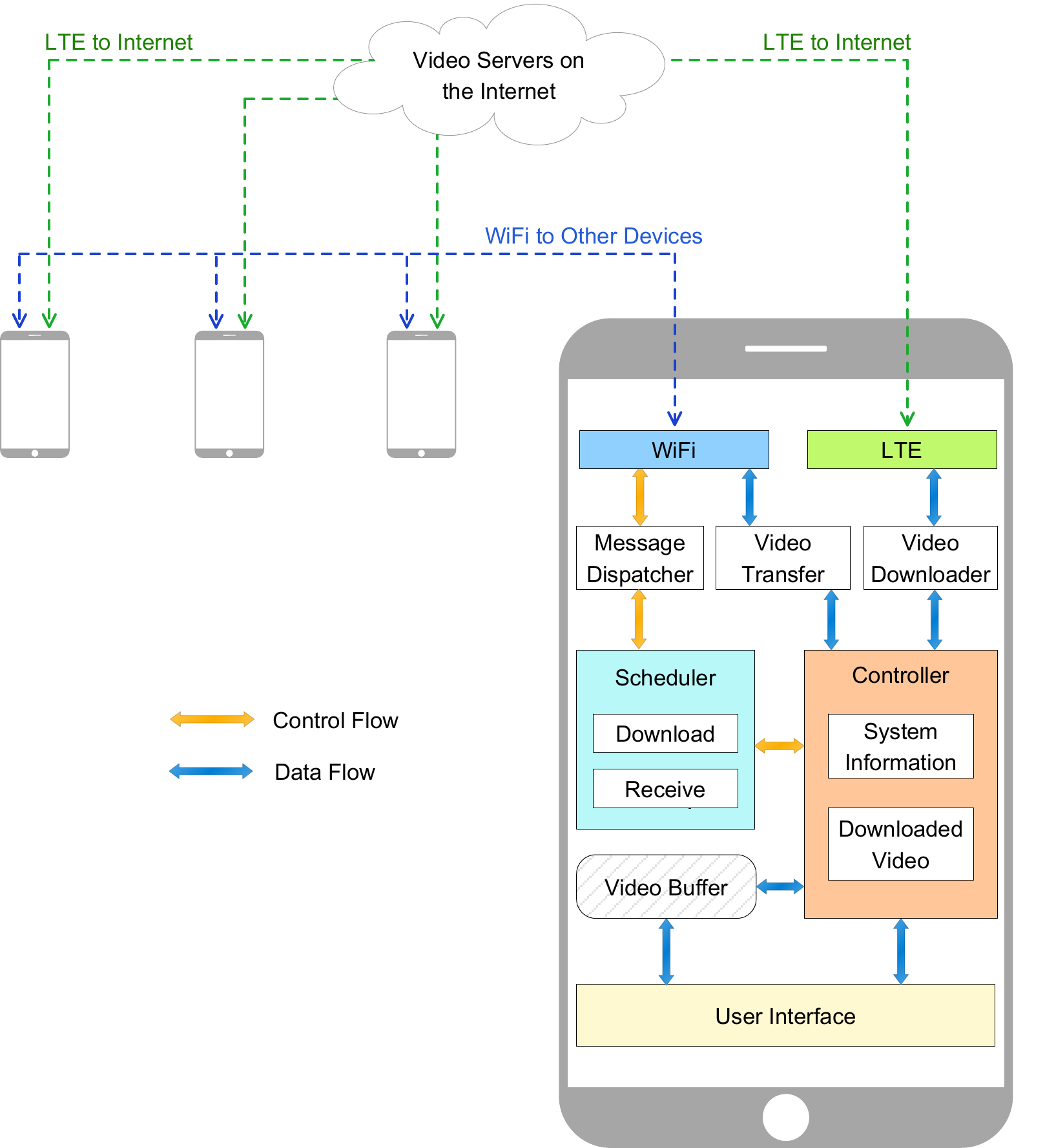}
  \caption{Demo System Architecture.}
  \label{fig:demo}
\end{figure}

\emph{1) Architecture:}
Figure \ref{fig:demo} shows the demo system architecture with
4 mobile devices (Raspberry PI devices), where mobile devices are connected with each other via WiFi  and connected to the video server on the Internet via LTE.
The demo system consists of the following (software) modules built in each mobile device.
The ``\textbf{Controller}'' module is the ``heart'' of the system and responsible for storing key information (such as system information and downloaded video data) and offering necessary control signal for other components.
The ``\textbf{Scheduler}'' module is another key component in the system and responsible for implementing our proposed online Lyapunov algorithm and making the scheduling decision.
It mainly consists of two components ``Download'' and ``Receive'': (i) when the device acts as a downloader helping others, the ``Download'' is active and in charge of the information announcement and scheduling determination;
and (ii) when the device acts as a receiver, the ``Receive''   is active and in charge of information submission.
The ``\textbf{Video Downloader}'' module downloads video segments from video servers on the Internet through LTE links.
The ``\textbf{Video Transfer}'' module is responsible for transmitting and receiving the downloaded video data among devices through WiFi links.
The ``\textbf{Message Dispatcher}'' module is responsible for transmitting and
receiving the control messages (such as buffer length, segment size, and url) among devices through WiFi links.
Finally, the ``\textbf{Video Buffer}'' module stores the segments that are for the user's own video consumption,
and the ``\textbf{User Interface}'' module fetches video segments from buffer and displays to users.

\emph{2) Operation:}
The demo system operates in the following way.
When a device is ready for downloading, its ``Message Dispatcher'' module starts to collect the necessary information of nearby users and pass the information to the ``Scheduler'' module.
Then, the ``Scheduler'' module makes the scheduling decision (i.e., for whom it is going to download the next segment) and passes the scheduling result, together with the necessary information (e.g., the segment id, encode bitrate, and url), to the ``Controller'' module.
Finally, the ``Controller'' module dispatches the ``Video Downloader'' module to download the video segment and the ``Video Transfer'' module to send the downloaded segment to the target device.

\emph{3) Signal Flow:}
The detailed signal flow is shown as follows.
When a device is ready to download a new segment (e.g., when completing a segment download), it initiates and broadcasts a ``READY'' message via WiFi.
Then, all nearby devices who receive the message and need helps will respond with the ``ACK'' message, together with the necessary context message (e.g., buffer length, segment size, encode bitrate, url, etc.), via WiFi.
Such an information exchange is performed by the ``Message Dispatcher'' module.
Next, when receiving the ``ACK'' messages from all nearby users, the device makes the scheduling decision by using the Lyapunov optimization framework in the ``Scheduler'' module, and then downloads the related video segment via LTE through the ``Video Downloader'' module.
Finally, the device passes the downloaded video data to the target device through the ``Video Transfer'' module.

\revaa{We notice that there may be many useless READY messages (without ACK replies), especially in the scenario where users are more likely to be in the idle mode than being in need of help (e.g., when users want to download video occasionally and do not have any download requests most of the time).
This may generate a lot of additional unnecessary overhead.
To reduce the unnecessary overhead caused by the unnecessary READY messages, we further introduce a ``sleeping mode'' for downloaders. The idea is to \emph{put downloaders to ``sleep'' (as far as collaboration is concerned\footnote{Note that the
downloader can still work on his own other tasks at the same time.}) when there is no further downloading request}.
This can be achieved by two functions called ``Sleep'' and ``Awake'' in the Controller module of downloaders.
The main task of ``Sleep'' is to turn a downloader into the sleeping mode when the downloader fails to receive an ACK message in a pre-specified time frame (e.g., 10 seconds) after initiating a READY message.
A downloader in the sleeping mode will no longer generate new READY message, until it is awaken.
The main task of ``Awake'' is to awake a sleeping downloader when the downloader overhears an ACK message (possibly to other downloaders).
Of course, each requester needs to initiate a virtual ACK message at the very beginning
to awake all potential downloaders. Obviously, with the above ``Sleep'' and ``Awake'' functions, our
current approach can effectively reduce the unnecessary READY messages in the scenario mentioned above.}

\begin{figure}[t]
	\center
	\includegraphics[width=3.5in]{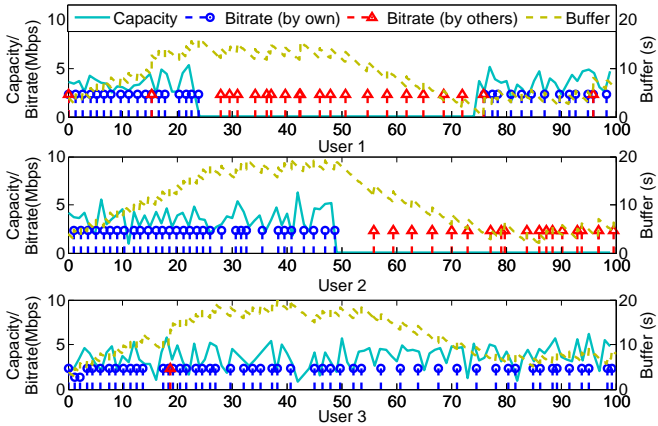}
		\caption{Cooperation between Connected and Disconnected Users.}
	\label{fig:demo-sim}
\end{figure}

\emph{4) Real Experiments:}
To illustrate the real performance of our proposed cooperative streaming, we further perform experiments over a demo system with 3 Raspberry PI devices, denoted by \{1, 2, 3\}.
The video server is set up on a lab computer, the video bitrates set is \{0.5, 1.0, 2.2, 5.0\}Mbps, and the segment length equals 10s.

We perform experiments in the scenario where devices have different cellular link capacities.
The goal is to illustrate how high capacity devices can help low capacity devices to improve the overall video streaming stability and performance of all devices.
In these experiments, each device has an average cellular link capacity of 3.5Mbps (when connecting to the Internet). Device 3 is always connected to the Internet, while device 1 is disconnected from  the Internet during the 25th to the 75th seconds and device 2 is disconnected from  the Internet during the 50th to the 100th seconds.

Figure \ref{fig:demo-sim} illustrates the video scheduling results for all devices \{1, 2, 3\} in a particular experiment round.
Here x-axis corresponds to the video streaming time horizon (of 100 second).
In each subfigure, the green curve denotes the real-time cellular link capacity of the corresponding device,
the yellow dash curve denotes the real-time buffer level of the corresponding device,
the blue stems with ``circle'' denote the segments downloaded by the   device itself, and the red stems with ``triangle'' denote the segments downloaded by other devices  through cooperation.
From Figure \ref{fig:demo-sim}, we can see that although
devices 1 and 2 are disconnected from the Internet half of the time,
they are still able to download and play the video smoothly
with the help of device 3.
By averaging the results over multiple experiment rounds, we   find that   our proposed cooperative streaming scheme can improve the average social welfare by up to 50.9\%  on average, comparing with the traditional non-cooperative streaming scheme.

\begin{figure*}[t]
\vspace{-5mm}
   \centering
  \includegraphics[width=.4\textwidth]{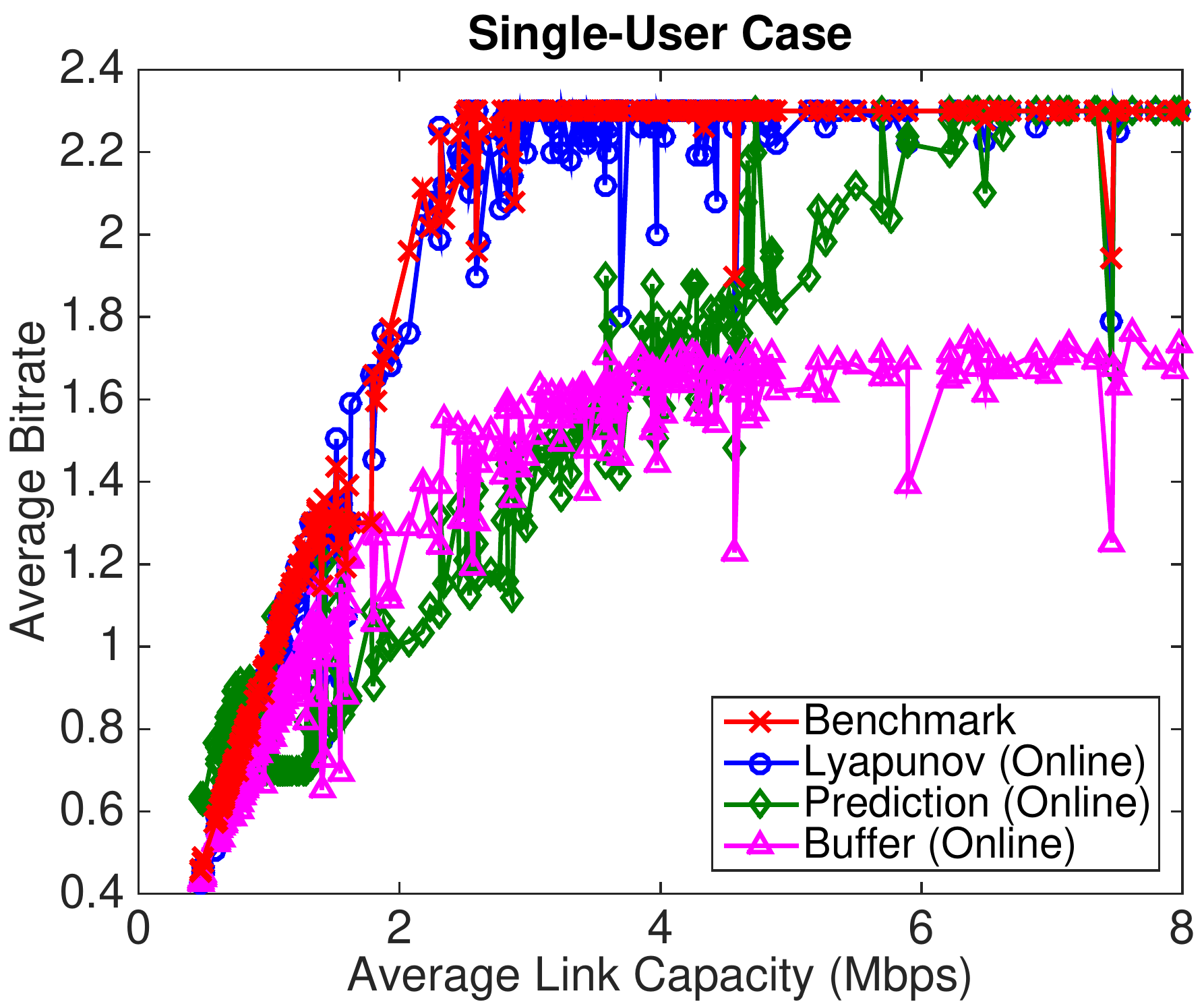}
  ~~~~~~~~~~~~~
  \includegraphics[width=.4\textwidth]{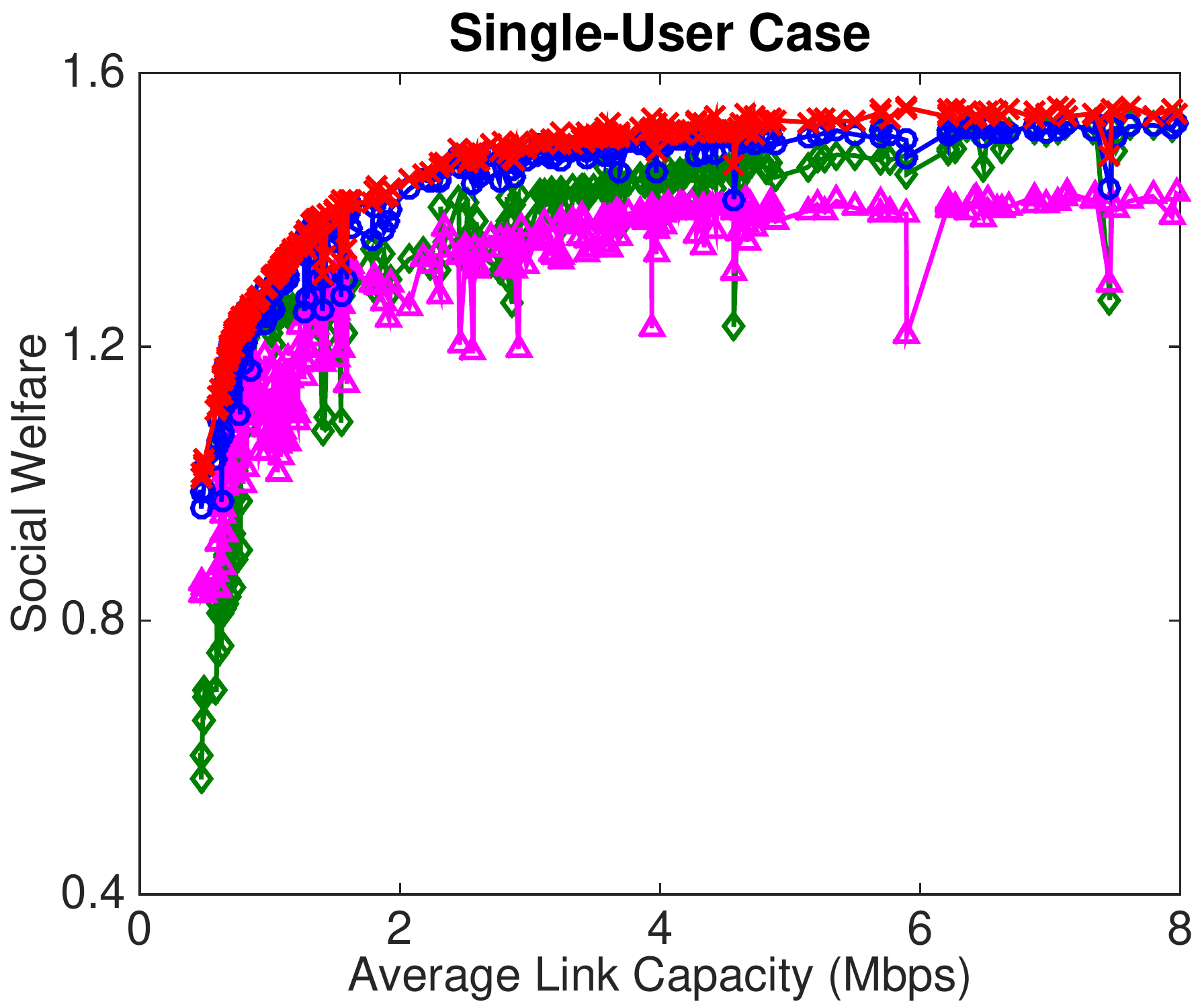}
  \vspace{-1mm}
  \\
~~~~
~~~~~~~~
~~~~~~~~
~~~~~~~~
(a)
~~~~~~~~
~~~~~~~~
~~~~~~~~
~~~~~~~~
~~~~~~~~
~~~~~~~~
~~~~~~~~
~~~~~~~~
(b)
~~~~~~~~
~~~~~~~~
~~~~~~~~
\\
\vspace{2mm}
  \includegraphics[width=.4\textwidth]{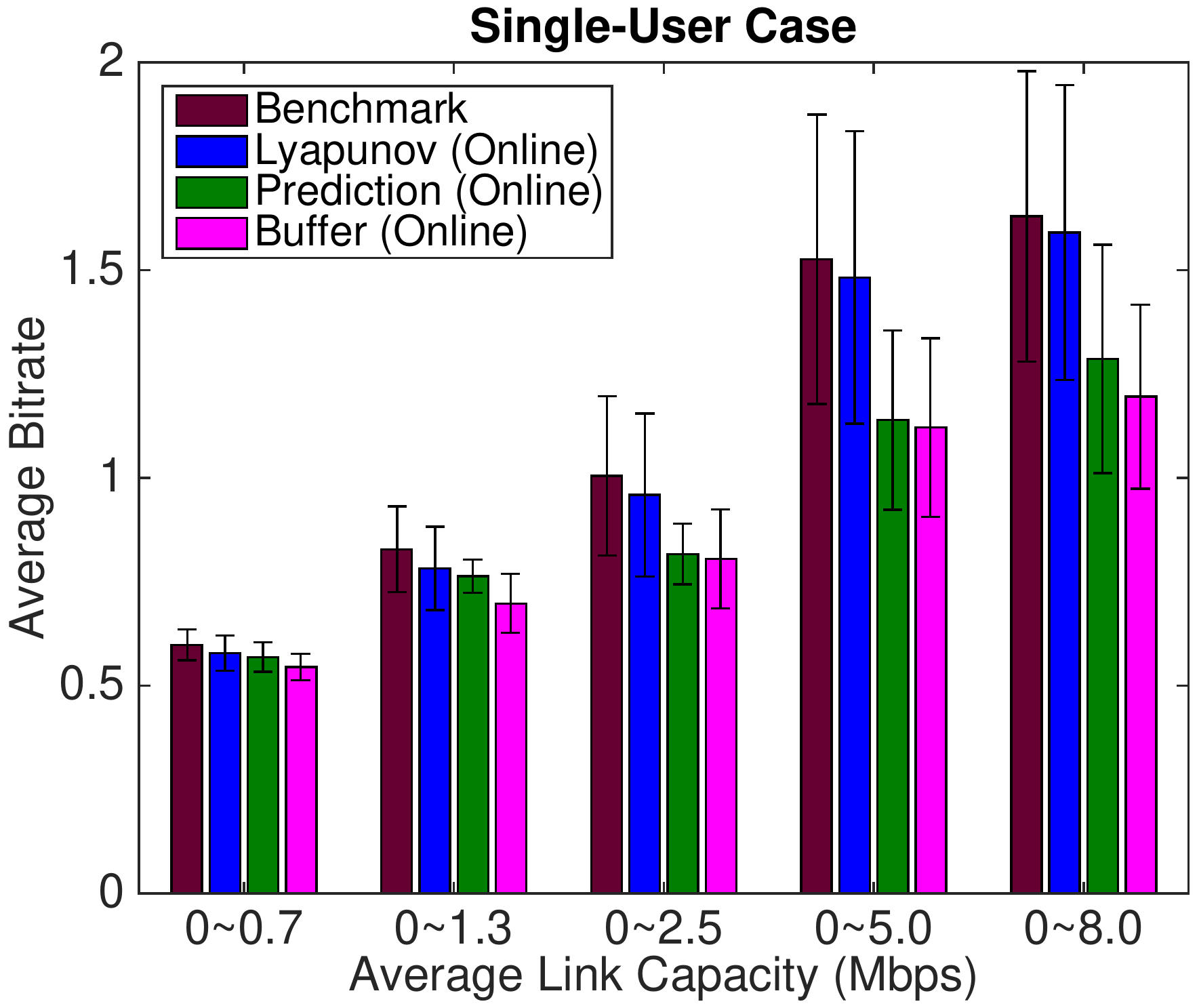}
  ~~~~~~~~~~~~~
  \includegraphics[width=.4\textwidth]{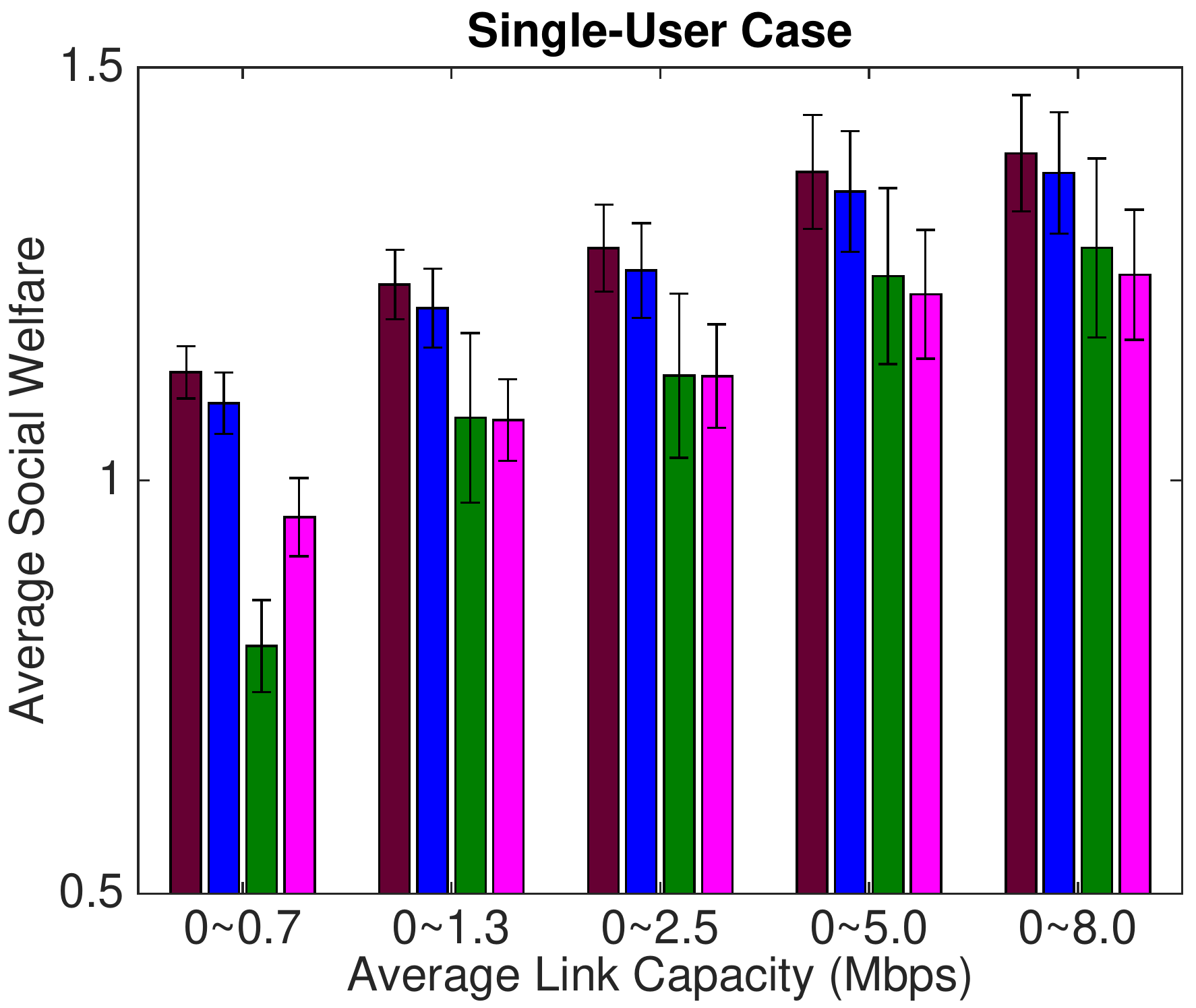}
  \vspace{-1mm}
\\
~~~~
~~~~~~~~
~~~~~~~~
~~~~~~~~
(c)
~~~~~~~~
~~~~~~~~
~~~~~~~~
~~~~~~~~
~~~~~~~~
~~~~~~~~
~~~~~~~~
~~~~~~~~
(d)
~~~~~~~~
~~~~~~~~
~~~~~~~~
  \caption{(a) Average Bitrate in Each Experiment, (b) Social Welfare in Each Experiment, (c) Average Bitrate in 1000 Experiments, (d) Average Social Welfare  in 1000 Experiments.
  (Red: Offline-Benchmark,
  Blue:  Lyapunov-Online,
  Green:   Prediction-Online,
  Pink:  Buffer-Online)}
  \label{fig:single-real}
\vspace{-3mm}
\end{figure*}

\subsection*{A.2 Single-User Simulation Results}

Now we construct experiments for the single-user scenario (i.e., non-cooperative scenario), where the user plays a high-resolution video (bitrate $2.3$Mbps). Similar as in the multi-user experiments, the total video length is 500 seconds,  the segment length is 2 seconds, and the maximum  buffer length at the user's device is 40 seconds.
 {We use these single-user experiments to illustrate the performance gap of our proposed Lyapunov-based online algorithm to the theoretical performance bound.
We also use these experiments to compare the bitrate adaptation performance of our proposed algorithm with the existing online algorithms.}


Figure \ref{fig:single-real} shows the average bitrate and social welfare under different average link capacities (extracted from the measured link throughput traces).
Red curve/black bar denotes the theoretical upperbound (benchmark), Blue curve/bar denotes the proposed Lyapunov-based online algorithm,
Green curve/bar denotes the  channel prediction-based algorithm in \cite{b6},
and Pink curve/bar denotes the
buffer-based  algorithm in \cite{b1}.
Each point in subfigures (a) and  (b) denotes the average bitrate and   social welfare generated in one experiment (corresponding to a particular choice of data trace), respectively.
Subfigures (c) and (d) shows the average bitrate and average social welfare in 1000 experiments under different average link capacity ranges.
For example, in the first bar group, we calculate the average bitrate and average social welfare achieved in all experiments with an average link capacity below 0.7Mbps.

We can see from (a) and (c) that our proposed algorithm (Blue) achieves
an average bitrate   higher than other two  algorithms (with an average bitrate increase of $5\%\sim30\%$), and is very close to the offline benchmark (Red).
We can further see from (b) and (d) that our proposed algorithm achieves an average social welfare higher than other two algorithms (with an average gain of $10\% \sim 40\%$), and is very close to the theoretical upperbound (with an average gap less than $3\%$).
Moreover, the social welfare gain decreases with the maximum link capacity.
This is because with a larger link capacity, all algorithms approach to the upperbound, hence their differences become less significant.

 {The above experiments demonstrate
that the bitrate adaptation mechanism in our   algorithm is better than those in \cite{b1} and \cite{b6}. By Theorem 2,
our algorithm asymptotically
converges to the theoretical performance upperbound (with
a controllable gap), while the other two algorithms represent some reasonable heuristics without a theoretical performance guarantee.}


\begin{thebibliography}{99}



\bibitem{CiscoReport} Cisco VNI: Global Mobile Data Traffic Forecast Update, 2016-2021.



\bibitem{abr}
 S. Akhshabi, Ali C. Begen, and C. Dovroli,
 ``An Experimental Evaluation of Rate-Adaptation Algorithms in Adaptive Streaming over HTTP,''  \emph{Proc. ACM MMSys}, 2011.


\bibitem{adobe}
Adobe Systems, ``HTTP Dynamic Streaming,''
url: \url{http://www.adobe.com/products/hds-dynamic-streaming.html}


\bibitem{apple}
R.P. Pantos, ``HTTP Live Streaming draft-pantos-http-live-streaming-13,'' Network Working Group, 2014, url: \url{http://tools.ietf.org/html/draft-pantos-http-live-streaming-13}


\bibitem{microsoft}
Microsoft, ``Smooth Streaming,'' url: \url{http://www.iis.net/downloads/microsoft/smooth-streaming}



%

%

%
%
%




\bibitem{b1} T. Y. Huang, R. Johari, N. McKeown, M. Trunnell, and M. Watson, ``A buffer-based approach to rate adaptation: Evidence from a large video streaming service," \emph{Proc. ACM SIGCOMM}, 2014.

\bibitem{b6} Z. Li, X. Zhu, J. Gahm, R. Pan, H. Hu, A. C. Begen, D. Oran, ``Probe and adapt: Rate adaptation for http video streaming at scale," \emph{IEEE Journal on Selected Areas in Communications}, 32(4):719-733, 2014.



%
%
%
%


%
%





\bibitem{ming-maga}
M. Tang, L. Gao, H. Pang, J. Huang, and L. Sun, ``Optimizations and Economics of Crowdsourced Mobile Streaming,'' \emph{IEEE Communications Magazine}, 55(4):21-27, 2017.



\bibitem{c6} Y. Zhang, C. Li, L. Sun, ``DECOMOD: collaborative DASH with download enhancing based on multiple mobile devices cooperation," \emph{Proc. ACM MMSys}, 2014.



\bibitem{add-1}
N. Golrezaei, P. Mansourifard, A. F. Molisch, and A. G. Dimakis,
``Base-Station Assisted Device-to-Device Communications for High-Throughput Wireless Video Networks,'' \emph{IEEE Transactions on Wireless Communications}, 13(7):3665-3676, 2014.




\bibitem{add-3}
L. Keller, A. Le, B. Cici, H. Seferoglu, C. Fragouli, and A. Markopoulou
``MicroCast: Cooperative Video Streaming on Smartphones,'' \emph{Proc. ACM MobiSys}, 2012.


\bibitem{add-3a}
Y. Cao, X. Chen, T. Jiang, and J. Zhang, ``SoCast: social ties based cooperative video multicast,'' \emph{Proc. IEEE INFOCOM}, 2014.



\bibitem{multiuser-dash-1}
W. Pu, Z. Zou, and C. W. Chen, ``Video adaptation proxy for wireless dynamic adaptive streaming over HTTP,'' \emph{IEEE Workshop Packet Video}, 2012.




\bibitem{ming-2017}
M. Tang, S. Wang,  L. Gao, J. Huang, and L. Sun, ``MOMD: A Multi-Object Multi-Dimensional Auction for Crowdsourced Mobile Video Streaming,'' \emph{Proc. IEEE INFOCOM}, 2017.

\bibitem{ming-2016a}
M. Tang, L. Gao, H. Pang, J. Huang, and L. Sun, ``Multi-Dimensional Auction Mechanism for Mobile Crowdsourced Video Streaming,''
\emph{Proc. IEEE WiOpt}, 2016.

\bibitem{ming-2016b}
L. Gao, M. Tang, H. Pang, J. Huang, and L. Sun, ``Performance Bound Analysis for Crowdsourced Mobile Video Streaming,''
\emph{Proc. IEEE CISS}, 2016.



\bibitem{c1x}
X. Kang and Y. Wu, ``Incentive Mechanism Design for Heterogeneous Peer-to-Peer Networks: A Stackelberg Game Approach,'' \emph{IEEE Transactions on Mobile Computing}, 14(5):1018-1030, 2015.

\bibitem{c2} M. Klusch, P. Kapahnke, X. Cao, B. Rainer, C. Timmerer, and S. Mangold,
    ``MyMedia: mobile semantic peer-to-peer video search and live streaming," \emph{Proc. ACM MOBIQUITOUS}, 2014.

\bibitem{c3} B. Rainer, C. Timmerer,  P. Kapahnke, and M. Klusch, ``Real-time multimedia streaming in unstructured peer-to-peer networks," \emph{Proc. IEEE CCNC}, 2014.




\bibitem{upn-lin}
G. Iosifidis, L. Gao, J. Huang, and L. Tassiulas, ``Incentive Mechanisms for User-Provided Networks," \emph{IEEE Communications Magazine}, 52(9):20-27, 2014.



\bibitem{opengarden-lin}
G. Iosifidis, L. Gao, J. Huang, and L. Tassiulas, ``Enabling Crowd-Sourced Mobile Internet Access," \emph{Proc. IEEE INFOCOM},  2014.


\bibitem{opengarden}
Open Garden, url: http://opengarden.com/


\bibitem{karma}
Karma, url: https://yourkarma.com/

\bibitem{att}
Tethering of AT\&T, url: www.att.com/shop/wireless/tethering.html



\bibitem{live-1}
K. Mori, S. Hatakeyama, H. Shigeno, ``DCLA: Distributed Chunk Loss Avoidance Method for
Cooperative Mobile Live Streaming,'' \emph{Proc. IEEE AINA}, 2015.

\bibitem{live-2}
T. Wu, W. Dou, Q. Ni, S. Yu, and G. Chen, ``Mobile Live Video Streaming Optimization via Crowdsourcing Brokerage,'' \emph{IEEE Transactions on Multimedia}, 19(10):2267-2281, 2017.


\bibitem{liveapp-1}
IngKee, url: www.ingkee.com

\bibitem{liveapp-2}
Youtube Live, url: www.youtube.com/live dashboard splash

\bibitem{liveapp-3}
Facebook Livestream, url: www.facebook.com/livestream




\bibitem{inc-3}
G. Iosifidis, L. Gao, J. Huang, and L. Tassiulas, ``A Double Auction Mechanism for Mobile Data Offloading Markets,'' \emph{IEEE/ACM Transactions on Networking}, 23(5):1634-1647, 2014.

\bibitem{inc-1}
T. Luo,  S. S. Kanhere, J. Huang, S. K. Das, and F. Wu, ``Sustainable Incentives for Mobile Crowdsensing: Auctions, Lotteries, and Trust and Reputation Systems,''
\emph{IEEE Communications Magazine}, 55(3):68-74, 2017.


\bibitem{inc-4}
L. Gao, Y. Xu, and X. Wang, ``MAP: Multi-Auctioneer Progressive Auction for Dynamic Spectrum Access,''
\emph{IEEE Transactions on Mobile Computing}, 10(8):1144-1161, 2011.

\bibitem{inc-2}
C. Jiang, L. Gao, L. Duan, and J. Huang, ``Data-Centric Mobile Crowdsensing,''
\emph{IEEE Transactions on Mobile Computing}, 2017.



\bibitem{cccc-1}
Q. Ma, L. Gao, Y.F. Liu, and J. Huang, ``Incentivizing Wi-Fi Network Crowdsourcing: A Contract Theoretic Approach,'' \emph{IEEE/ACM Transactions on Networking}, 2018.

\bibitem{cccc-2}
L. Gao, J. Huang, Y. Chen, and B. Shou, ``An Integrated Contract and Auction Design for Secondary Spectrum Trading,'' \emph{IEEE Journal on Selected Areas in Communications}, 31(3):581-592,  2013.

\bibitem{cccc-4}
L. Duan, L. Gao, and J. Huang, ``Cooperative Spectrum Sharing: A Contract-based Approach,''
\emph{IEEE Transactions on Mobile Computing}, 13(1):174-187, 2014.

\bibitem{cccc-3}
L. Gao, X. Wang, Y. Xu, and Q. Zhang, ``Spectrum Trading in Cognitive Radio Networks: A Contract-Theoretic Modeling Approach,''
\emph{IEEE Journal on Selected Areas in Communications}, 29(4):843-855,   2011.



\bibitem{inc-5}
X. Zhang,  Z. Yang,  W. Sun,  Y. Liu,  S. Tang,  K. Xing, and  X. Mao, ``Incentives for mobile crowd sensing: A survey,'' \emph{IEEE Communications Surveys \& Tutorials}, 18(1):54-67, 2016.

\bibitem{inc-8}
L. Gao, G. Iosifidis, J. Huang, L. Tassiulas, and D. Li, ``Bargaining-based Mobile Data Offloading,''
\emph{IEEE Journal on Selected Areas in Communications}, 32(6):1114-1125,  2014.

%



%
%
%


%
%
%
%
%


%

\bibitem{mobility}
T. Camp, J. Boleng, and V. Davies, ``A survey of mobility models for ad hoc network research,''
\emph{Wireless Communications and Mobile Computing}, 2(5):483-502, 2002.


\bibitem{mobi-hotspot}
P. Yuan and H.-D. Ma, ``Opportunistic Forwarding with
Hotspot Entropy,'' \emph{Proc. IEEE WoWMoM}, 2013.


\bibitem{energy}
N. Balasubramanian, A. Balasubramanian, and A. Venkataramani, ``Energy consumption in mobile phones: a measurement study and implications for network applications,''
\emph{Proc. ACM SIGCOMM}, 2009.



\bibitem{energy-2}
M. A. Hoque, M. Siekkinen, and J. K. Nurminen, ``Energy efficient multimedia streaming to mobile devices¡ªa survey,''
\emph{IEEE Communications Surveys \& Tutorials}, 16(1):579-597, 2014.



\bibitem{lya}
 M. J. Neely, \emph{Stochastic Network Optimization with Application to
Communication and Queueing Systems}, Morgan \& Claypool, 2010.



\bibitem{smartphone-energy}
G. P. Perrucci, F. H. Fitzek, and J. Widmer, ``Survey on energy consumption entities on the smartphone platform,
\emph{Proc. IEEE VTC-Spring}, 2011.

\bibitem{data1}
\url{http://crawdad.cs.dartmouth.edu/ilesansfil/wifidog/}

\bibitem{data2}
\url{http://www.nextwifi.cn/wifind/}

\bibitem{data3}
\url{http://www.bestv.com.cn/}

\bibitem{report}
{Technical Report at arXiv}, url: \url{https://arxiv.org/abs/xxxx.xxxx}

\end{thebibliography}
\end{document}